\documentclass[runningheads]{llncs}
\usepackage{amsmath,amssymb,amsfonts}
\usepackage{hyperref}
\usepackage{bbm}
\usepackage{graphicx}
\graphicspath{{../pdf/}{../jpeg/}{figures}{authors}}
\usepackage{textcomp}
\usepackage{xcolor}
\usepackage{subfigure}
\usepackage{booktabs}
\usepackage{makecell}
\usepackage{bm}
\usepackage{enumitem}
\usepackage{setspace}
\usepackage{multirow}
\usepackage{longtable}
\usepackage{supertabular,booktabs}
\usepackage{geometry}
\begin{document}
\title{No driver, No Regulation? —— Online Legal Driving Behavior Monitoring for Self-driving Vehicles
}
%
%
\author{Wenhao Yu\inst{1,\dag}\and
Chengxiang Zhao\inst{2,\dag}\and
Jiaxin Liu\inst{1}\and
Yingkai Yang\inst{1}\and
Xiaohan Ma\inst{2}\and
Jun li\inst{1}\and
Weida Wang\inst{2}\and
Hong Wang\inst{1,*}\and
Xiaosong Hu\inst{3,*}\and
Ding Zhao\inst{4}
}
\authorrunning{WH. Yu et al.}
%
\institute{School of Vehicle and Mobility, Tsinghua University, Beijing, 100084, China 
\email{hong\_wang@tsinghua.edu.cn} \and
School of Mechanical Engineering, Beijing Institute of Technology, Beijing, 100081, China \and
Department of Mechanical and Vehicle Engineering, Chongqing University, Chongqing, 400044, China
\email{xiaosonghu@ieee.org}\and
Department of Mechanical Engineering, Carnegie Mellon University, Pittsburgh, PA, 15213, USA
}
\maketitle              
\begin{abstract}
Defined traffic laws must be respected by all vehicles.
However, it is essential to know which behaviors violate the current laws, especially when a responsibility issue is involved in an accident.
This brings challenges of digitizing human-driver-oriented traffic laws and monitoring vehicles’ behaviors continuously. To address these challenges, this paper aims to digitize traffic law comprehensively and provide an application for online monitoring of legal driving behavior for autonomous vehicles.
This paper introduces a layered trigger domain-based traffic law digitization architecture with digitization-classified discussions and detailed atomic propositions for online monitoring. The principal laws on a highway and at an intersection are taken as examples, and the corresponding logic and atomic propositions are introduced in detail.
Finally, the digitized traffic laws are verified on the Chinese highway and intersection datasets, and defined thresholds are further discussed according to the driving behaviors in the considered dataset. 
This study can help manufacturers and the government in defining specifications and laws and can also be used as a useful reference in traffic laws compliance decision-making. Source code is available on \url{ https://github.com/SOTIF-AVLab/DOTL}.

\end{abstract}
\section{Introduction}
Current traffic laws represent relatively stable driving regulations followed by the majority of drivers, which is essential for ensuring driving safety.Meanwhile, monitoring a vehicle behavior’s law compliance can provide substantial evidence for the traceability of traffic accidents. 
Due to the rapid development of autonomous vehicles (AVs), in the foreseeable future, there is going to be a period when AVs and human drivers will drive on the road together \cite{nair2021sharing}. This requires AVs to follow the traffic laws strictly in the same way as human drivers follow them\cite{cascetta2022autonomous,quante2021human}; otherwise, differences between humans and AVs' driving behaviors will lead to misunderstanding and distrust between humans and AVs, leading to chaos in the traffic flow and severely reducing driving safety.

However, how to make AVs follow the laws has been a challenge because the systematic solutions to traffic law compliance definitions and decision-making have still been under slow progress\cite{li2020deep,shetty2021safety,liu2022road}. Defining which behaviors comply with traffic laws represents the first step towards achieving law-abiding driving by AVs. However, this task remains challenging due to the fuzziness inherent in natural language traffic laws, which are oriented towards human drivers. Given the current technical limitations, it is difficult for AVs to understand natural language, especially when dealing with safety-critical laws that rely on human knowledge. Specifically, AVs can only interact with digital information that has precise meanings, which raises the question of how to accurately express the current fuzzy natural language traffic laws in digital form.
\begin{figure}[thbp]
    \centering
    \includegraphics[width = 1\linewidth]{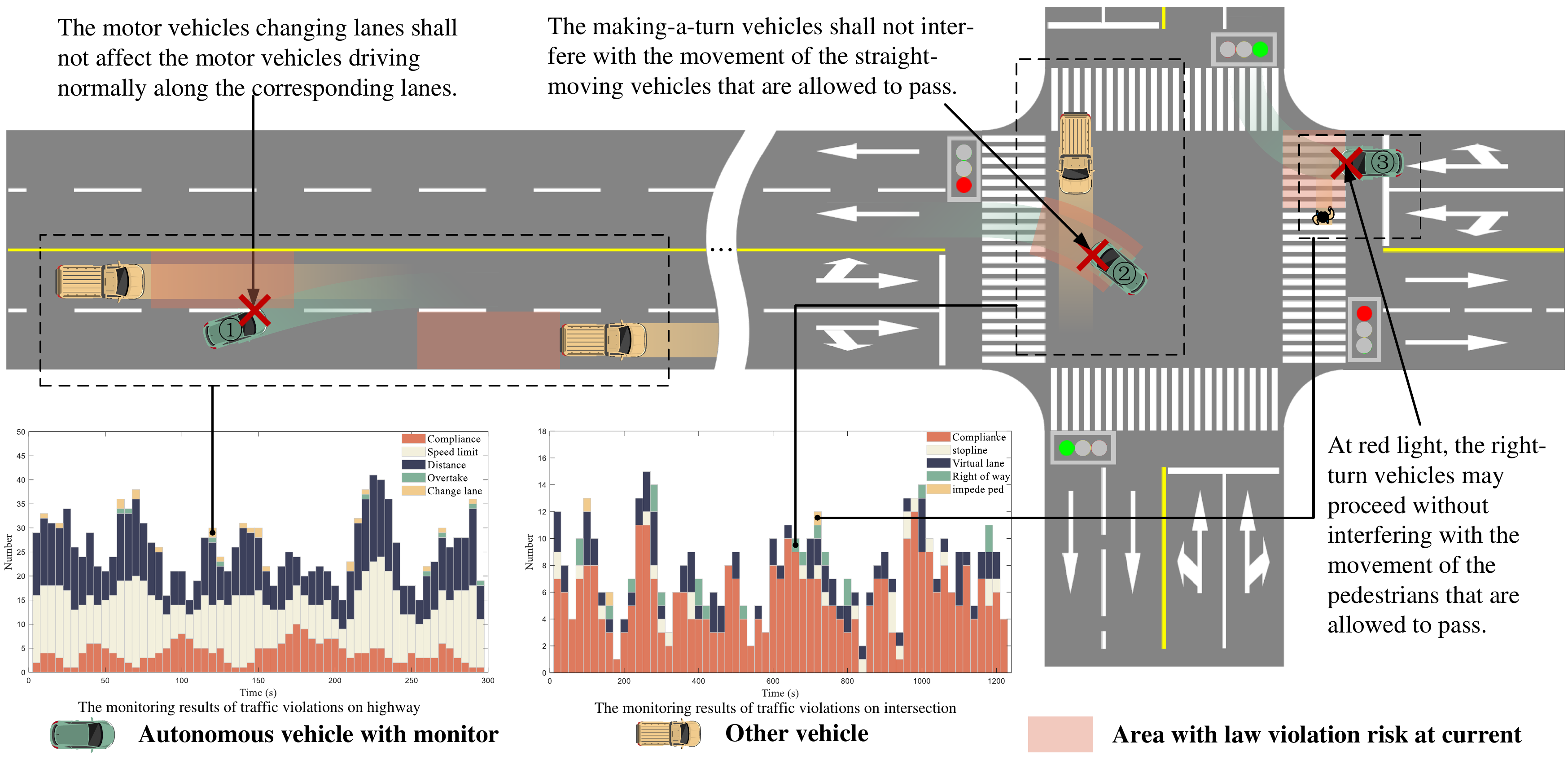}
    \caption{Online traffic law violation monitoring of autonomous vehicle.This monitor can be deployed on the ego vehicle and monitoring the traffic law violation behaviors of the ego vehicle in both highway and intersection scenarios. When deployed on every vehicle in the dataset, the statistical chart of monitoring results are shown as the two figure at the left lower corner.}
    \label{violation monitor}
\end{figure}

In complex traffic scenarios, AVs not only need to avoid pedestrians, motor vehicles, and other traffic participants but also must follow traffic law constraints, such as traffic signs, traffic markings, and right-of-way laws. However, current natural language traffic laws cannot be directly transformed into executable automatic driving commands. Compliance understanding of fuzzy natural language in research on traffic laws varies from person to person, making it challenging to map traffic law sentences into static, fixed compliance logic judgment expressions. Therefore, constructing a standardized mathematical description language of traffic laws that covers individual understanding differences and complex traffic law constraints is crucial for breaking through the bottleneck of intelligent vehicle-compliant driving technology.

\cite{buchanan1970some} conducted the systematical law standardisation research for the first time using expert systems and a rule-based standardised approach.
Sergot et al. and Bench-Capon et al. focused on the standardization of laws using first-order logic statements\cite{1986The,1987Logic}. Their works represent significant milestones in the development of standardized languages for describing laws, which can be applied to various domains, including traffic rules.
However, their research has been limited in expressing the original concepts. In addition to the first-order logic, there has been a deontic logic-based method for standardising the rules, which can explicitly express the notions of permission and obligation\cite{1983Permissions,governatori2018practical}.
\cite{bhuiyan2019methodology,villata2020traffic} proposed a defeasible deontic logic to handle rule exceptions and resolve conflicts in rule norms. For road traffic laws, \cite{Royakkers0Extending} extended the obligation logic to solve the problem of speed limit conflict in Dutch traffic laws. In addition to these methods, most studies have identified temporal logic as a suitable specification formalism. Linear temporal logic (LTL) \cite{pnueli1977temporal} is a specification formalism based on propositional logic and temporal operators, which has been used to specify motions and formalize traffic laws in most studies, including those by Esterle et al. \cite{esterle2020formalizing,esterle2019specifications}. Further, metric temporal logic (MTL) represents an extension of propositional linear temporal logic with discrete-time-bounded temporal operators~\cite{thati2005monitoring}. Most commonly occurring real-time properties, such as invariance and bounded response, can be expressed in MTL fragments~\cite{ouaknine2008some}. The MTL has been used for traffic laws monitoring because it can specify an interval over which a particular property must be fulfilled~\cite{maierhofer2020formalization}. Unlike LTL and MTL that provide Boolean results, the result of signal temporal logic (STL)~\cite{maler2004monitoring} with quantitative semantics \cite{fainekos2009robustness} represents a degree of satisfaction or violation of a property \cite{hekmatnejad2019encoding,arechiga2019specifying,fremont2022scenic}. 

Based on the research on logic expressions of traffic laws discussed earlier, \cite{rizaldi2015formalising,rizaldi2016formally,rizaldi2017formalising}decomposed traffic laws into a set of atomic propositions and used LTL to express logic combinations using these atomic propositions. By employing the Isabelle theorem, they were able to achieve a standardized description of traffic laws. The proposed method was validated on a dataset by monitoring overtaking and safe distance. However, this method can only perform offline monitoring based on data recorded by a vehicle and cannot be applied in real-time during actual vehicle driving. To achieve real-time online monitoring, \cite{esterle2020formalizing} used LTL to describe traffic laws and transformed these laws into a deterministic finite automaton to monitor the compliance of vehicles' behaviors and combined them with datasets to eliminate misjudgments iteratively. However, this method is applicable only to laws of interactive behaviors of road participants and does not consider the rules for the representation of static participants, such as traffic signs and markings. \cite{9488998} proposed an online monitoring method but from a third-party view. They used Mask R-CNN to distinguish moving objects first and then estimated their parameters; next, binary thresholds were used to judge whether the object broke the laws. This method is faster and more efficient than humans but can be applied only to detecting traffic marking law violations and can be used only in roadside equipment.
Roadside video monitoring methods have limitations in terms of their ability to directly contribute to autonomous driving, as they typically rely on external monitoring systems rather than on-board sensors and algorithms. However, they can provide valuable insights and data for the development of autonomous driving systems, particularly in the areas of law compliance and behavior analysis \cite{wang2013video,csentacs2019real,bachtiar2020vehicle}.
In terms of vehicle side monitoring and simultaneous monitoring of static participant laws and interactive behaviors, \cite{esterle2019specifications} defined vehicle behavior through discrete spatial sequence combinations, extended monitoring to decision trajectories, and judged the compliance of the action sequence of each trajectory. However, this method requires road space division based on other vehicles and environmental information, which is time-consuming in complex scenes and can reduce accuracy. Further, \cite{ozkul2018police} proposed a vehicle side but not Ego vehicle online monitor system, which uses communication with nearby vehicles to monitor other vehicles' law-violation behaviors and reports the detected behaviors to the transportation authority. This system preserves privacy and has no false positives, but the original study paid more attention to introducing the proposed framework while fewer law violation monitor algorithms were discussed. The judgement method proposed in \cite{karimi2020formalizing} for right-of-way monitoring is effective in ensuring safety, as it requires vehicles to follow the rules strictly and to use turn signals correctly. However, this approach might be too conservative and result in excessive yielding and traffic congestion in certain scenarios. Therefore, a more balanced approach is necessary to ensure safety while also optimizing traffic flow. 

The above-mentioned research has provided a useful theoretical basis for the digitization of traffic laws and the monitoring of law violations by high-level AVs. However, standardized and systematic mathematical descriptions of traffic laws and comprehensive violation monitoring that incorporate different types of law constraints remain serious challenges. It can be observed that current standardized or formalized descriptions of traffic laws are only in the stage of trying different logical expressions to digitize some law articles. However, there is no complete scheme for online monitoring of traffic law violations of ego vehicle behavior at the vehicle end. This paper proposes a systematic method to monitor self-traffic law violation behaviors from the ego vehicle side, as shown in Fig.\ref{violation monitor}. By applying this method to each car in the highway and intersection dataset, all violation behaviors can be found and counted as the two figures in the lower left quarter of Fig.\ref{violation monitor}.

The contributions of this paper are as follows:

1) An ego vehicle side traffic law violation online monitor system that only use perception and decision-making information is proposed, whose result can be used for better decision-making. A traffic law constraints classifications and online monitor-oriented trigger domain-based layered architecture for traffic law digitization is established.

2) The detailed computer-friendly atomic proposition list involved traffic law digitization and the example MTL expressions for primary highway and intersection laws is established.

3) The essential thresholds selection method and the dataset verification method  are proposed, with its results of the example online monitor.

\section{Traffic laws and Constraints}

Various countries and regions around the world have their own distinct road traffic laws, shaped by local culture, history, and social background. Nevertheless, these laws share consistency in their fundamental principles, all geared towards ensuring the safety of both drivers and pedestrians. Moreover, there exists a significant degree of similarity in the constraints imposed on driving behavior. For instance, most countries and regions worldwide have established maximum speed limits on roads and require drivers to maintain sufficient following distance with the vehicle ahead. With respect to right of way, the majority of regions adopt the principle of priority, albeit with slight variations in the allocation of right-of-way. Moreover, the expressions and meanings of traffic signs and markings in different regions are fundamentally alike. These consistencies reflect the unite of the internal logic of different countries' traffic laws to guaranteeing road safety.

Although there exist variations in laws among different countries, these differences primarily lie in the threshold of constraints on different behaviors. However, the behaviors that are constrained in the traffic laws and the means of constraint are generally consistent. Above consistency makes it possible to use one digital framework of laws to solve the digitization problem of different laws in different regions. The road traffic laws in most countries and regions mainly limit driving behaviors from the following four aspects: vehicle speed, distance, actions, and road rights. 
 In the subsequent sections of this article, we will employ China's relevant road traffic laws, namely, \textit{The Regulation on the Implementation of the Law of the People's Republic of China on Road Traffic Safety}, as an example to illustrate our work.

\begin{figure}[htbp]
    \centering
    \includegraphics[width = 0.4\linewidth]{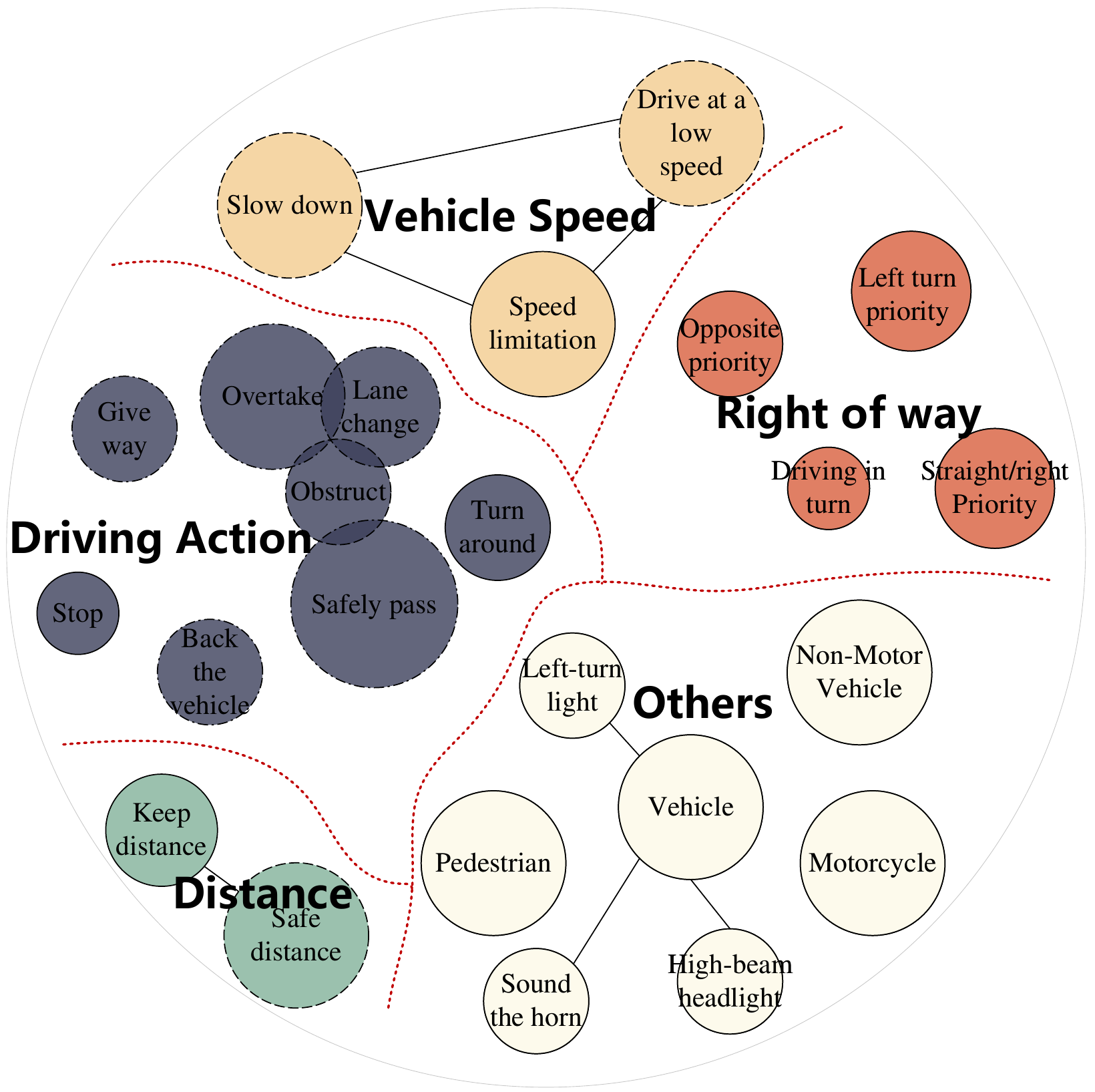}
    \caption{The classifications of Chinese traffic laws restrictions. Each circle means a different class of traffic law constraint. The size of the circle represents the proportion of this class, and the overlap means both constraint classes appear in the same article. The solid edge linetype of the circle meanings there is no ambiguous expression involved, while dash dot line means in certain articles fuzzy parameter is involved and the dot line means the corresponding expressions in each article need to clarify.}
    \label{classification}
\end{figure}

According to \textit{The Regulation on the Implementation of the Law of the People's Republic of China on Road Traffic Safety}, Chapter 4: Road traffic regulations, there are 49 traffic regulations, but only 25 articles are related to motor vehicle driving behaviors. Due to the overlapping, there are 11 articles restricting the vehicle speed; 3 articles restrict the distance between vehicles, 12 articles regulate driving behavior and four articles concern right of way. A detailed classification is shown in Fig.\ref{classification}.

Among all vehicle speed constraints, the traffic law usually restricts driving behavior through three types of descriptions: 1) defining upper or lower speed limits; 2) giving the driving behavior suggestions; 3) giving driving state suggestions. In distance-related constraints, the distance between vehicles is usually constrained by descriptions such as: 1) maintaining a specific distance in certain scenes, or 2) to maintain a safe distance. As for the constraints of actions, eight types of actions are regulated. The law specifies whether a vehicle shall pass, not hinder, overtake, change lanes, give way, turn around, reverse or stop under certain working conditions. Among all eight types of constraints, only the “not hinder” is fuzzy and without any further definitions. The “stop” and “turn around” actions are clearly define while other constraints are defined clearly only in certain situations. As for the constraints regarding the right of way, it stipulates when a vehicle has the right of way priority in a specific situation. for instance, the straight-forward or right-forward vehicles go first, left-turning vehicles go first, or driving in proper order.

Therefore, the Chinese traffic law and its regulations restrict vehicles’ driving behaviors under certain situations by direct or fuzzy natural language-based descriptions. Due to the fuzzy part of the descriptions, certain driving actions vary with drivers’ experiences, ages, risk tolerances. However, it is hard for machines to understand so many regulations, especially fuzzy ones. Therefore, it is necessary to digitize traffic laws using the same standard for all AVs based on the exact and executable logic descriptions.

\section{Traffic Law digitization}
\begin{figure}[ht!]
    \centering
    \includegraphics[width = 0.8\linewidth]{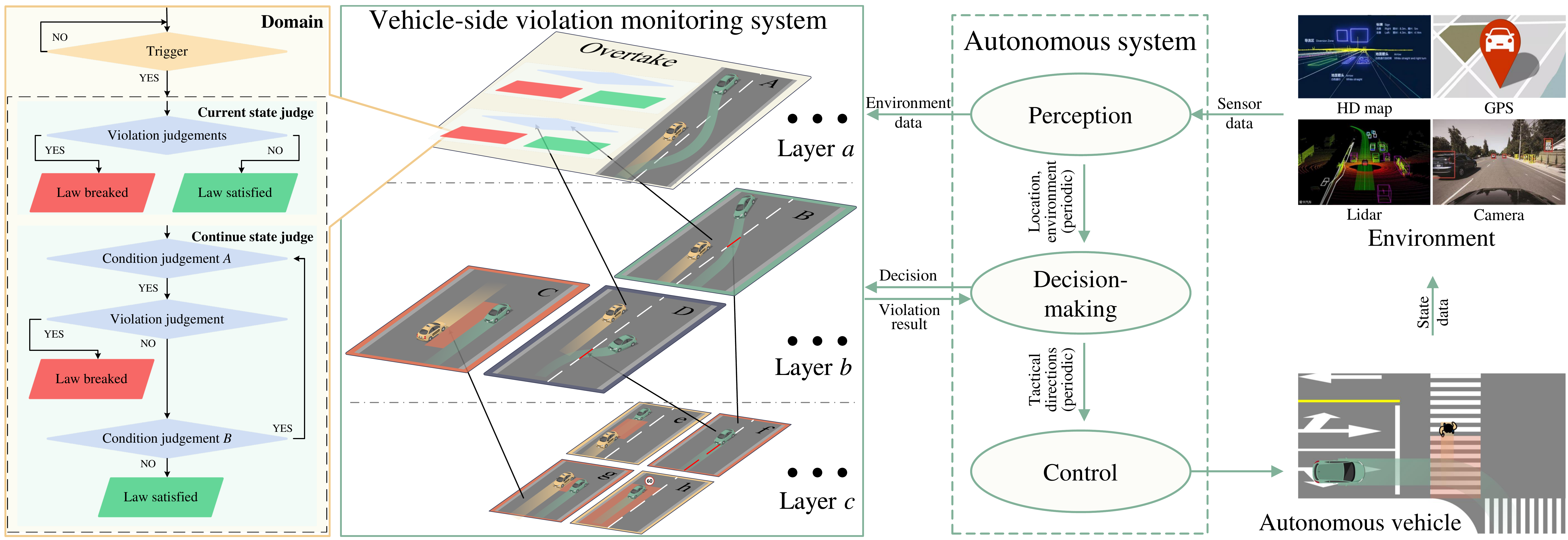}
    \caption{The architecture of the proposed trigger domain-based layered traffic law digitization method}
    \label{architecture}
\end{figure}
\subsection{Different purposes of law digitization}
When digitizing the traffic laws, the purpose of digitization will greatly influence the digitization progress and its difficulty. The purposes of traffic law digitization can be roughly divided into two main categories: offline monitoring oriented and online monitoring oriented. Offline monitoring is more like a posterior approach, as it can obtain the vehicle's behavior information for the entire period, making it easy to get the vehicle's next action during operation and selectively choose the regulations to be monitored. However, offline monitoring cannot serve as a real-time vehicle-side monitoring system. Online monitoring can only obtain current and historical information, and future information can only rely on prediction algorithms. Online monitoring is divided into three types based on the information used for monitoring: 1) fact-based law violation monitoring; 2) decision-based law violation monitoring;  3) prediction-based violation monitoring. 

Fact-based monitoring uses historical and current vehicle state data, but has limitations in determining vehicle behavior intentions (e.g., distinguishing between a vehicle's intention to simply change lanes to the right versus overtaking on the right), resulting in complex monitoring progress and greater difficulty in formulation. When the ego vehicle’s decision data are involved, online monitoring gains the ability to foresee the future actions of the ego vehicle, and this is referred to as decision-based law violation monitoring. This monitoring type can tell whether the ego vehicle will violate one non-interactive traffic law if following the current decision. Therefor, the decision-making system can benefit from reading the monitor’s output to adjust its decision to comply with the traffic law. Furthermore, prediction-based law violation monitoring considers the prediction behaviors of other participants. This monitoring type uses the predictive information to judge the law compliance of the decision trajectory in the prediction range,which reduce the condition classification discussions and easier to realize. Therefor, it is decision-friendly and gives the best advance quantity to adjust the decision. However, due to the reliance on prediction algorithms, the monitoring result is unstable and varies with the decisions, other participants’ behaviors and their predictions. 

It should be noted that online monitor can be installed on the vehicle, differ from the roadside equipment, its monitoring scope covers all vehicle running progress.
Moreover, besides the onboard law violation records, its ability to distinguish law violation behaviors can provide suggestions for every decision that an autonomous system makes. Therefore, according to the aforementioned analyses and reasons, this study selects the decision-based online law violation monitoring as the research objective, which is a balance between the difficulty and result meanings of the decision-making system.

\subsection{Trigger domain-based layered architecture of traffic law digitization}
Although traffic laws usually define different constraints under certain conditions, much research on law digitization and monitoring has been conducted for improving the offline methods. In the related research, conditions have usually been simplified for digitized laws because the scenario classifications or certain behavior monitoring groups imply the conditions that the digitized violation monitoring will confront with. Different from offline monitoring, in online monitoring, scenarios appear randomly. Thus, it is necessary to classify the current situation, select the right law article, and monitor the vehicle’s behavior. Therefore, condition classification plays an essential role in online monitoring. In view of that, this paper proposes a trigger domain-based layered architecture for law digitization. The proposed architecture is described in the following.

Fig.~\ref{architecture} depicts an online decision-based law violation monitor deployed on the ego vehicle. The monitor reads data from the perception and decision-making systems and outputs results to the decision-making system to aid compliance decision-making. To simplify the digitization process, certain law articles can be broken down into simple ones since some articles specify a series of vehicle actions. As a result, the laws can be grouped and digitized into layers, where the upper layer includes more articles than the lower layer. The proposed architecture avoids monitoring the same action multiple times since upper-layer articles can call the lower layer's digital expressions, and the lowest article code will run only once throughout the monitoring process. Each article's digital expression constitutes a domain, where the trigger serves as the domain's entrance. Only when a trigger is satisfied, the data flow enters the domain, and the violation judgement begins. These triggers are logical results of a series of logical judgements that act as condition classifiers to activate the corresponding law article for monitor under the right conditions. As it is an online monitor, it can only access past and present data. The violation judgements within the domain can be broadly classified into two types: current state judgements and continuous state judgements. Current state judgements can determine if the vehicle is violating a specific law or not, while continuous state judgement can determine if the vehicle's behavior illegal over a period of time or a cumulative state meets the criteria. This proposed architecture facilitates the logical digitization of traffic laws and ensures the monitoring process's accuracy and consistence.

\subsection{Atomic proposition for application}
Temporal logic has been widely used in most studies on traffic laws digitization and monitoring. As the proposed monitor contains continuous state judgement, we used metric temporal logic (MTL) in this study to interpret the definition of the trigger domain and logic proposition. MTL introduces more temporal operators and has good performance in describing the relationship between behavioral logic within a limited time and spatial coordinates.

 Given a set $\mathcal{AP}$ of atomic propositions, where each atomic proposition $\sigma_{i} \in \mathcal{AP}$ represents a Boolean statement, a MTL formula $\varphi$ is defined as
\begin{equation}
    \begin{aligned}
    \varphi ::= &~\textit{\textbf{T}}~| \sigma_{i}| \lnot\varphi| \varphi_{1}\land\varphi_{2}| \varphi_{1}\lor\varphi_{2}| \varphi_{1}\Leftrightarrow\varphi_{2}| \\
    &G_{I}(\varphi)|F_{I}(\varphi)|P_{I}(\varphi)|O_{I}(\varphi)| \varphi_{1}U\varphi_{2}
    \end{aligned}
\end{equation}
Where $\textit{\textbf{T}}$, $\lnot$, $\land$ and $\lor$ are the Boolean operators, a $\Leftrightarrow$ b means that a is equivalent to b. \textit{G}, \textit{F}, \textit{P} and \textit{O} are temporal operators. The subscript \textit{I} represents an interval $\mathbb{R}_{\geq 0}$ expressing time constraints relative to the current time. The globally operator \textit{G} indicates that $\varphi$ holds throughout the entire time sequence, and the future operator \textit{F} indicates that $\varphi$ holds within a time interval for some future state. The previously operator \textit{P} expresses that $\varphi$ holds within a time interval for the previous state, and the once operator \textit{O} specifies that $\varphi$ holds only once before a certain point in time. The Until operator \textit{U} specifies that a property holds true until another property becomes true.

\begin{figure*}[htbp]
    \centering
    \includegraphics[width = 0.8\linewidth]{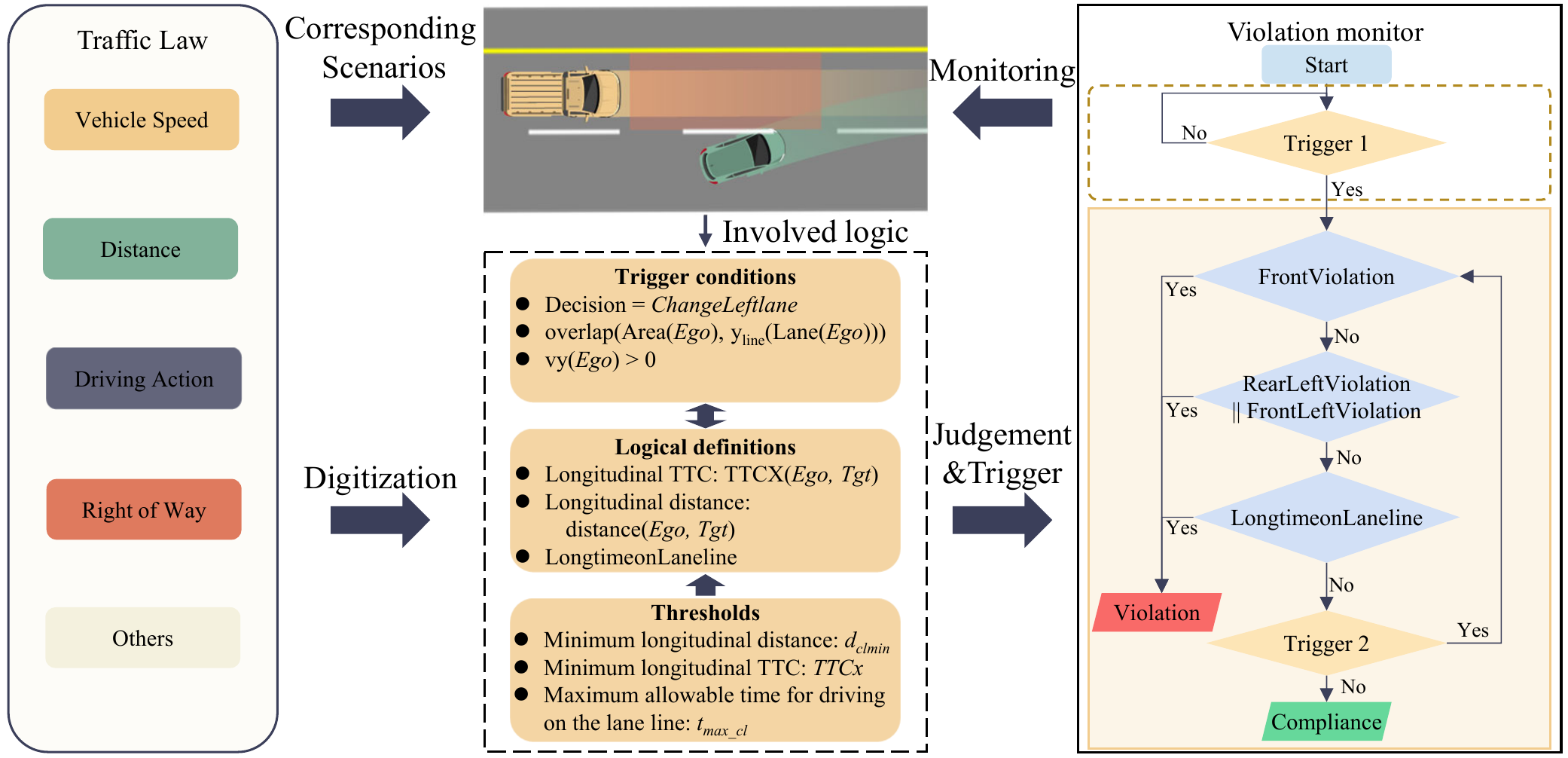}
    \caption{The overall process of traffic law digitization}
    \label{overallthinking}
\end{figure*}

Research on digitizing traffic laws using MTL expressions typically focuses on showing the logical relationships between atomic propositions. While logical relationships are important, they alone cannot provide the expression's results. The atomic propositions between logical operation symbols also contribute significantly to MTL expressions. Executing MTL expressions in programs requires using both logic relationships and calculable atomic propositions. However, most studies have not decomposed laws into the smallest calculable atomic propositions, making MTL expressions resemble natural language law formulation but in a logical combination way. To improve the traffic law monitor's practicality, this study decomposes each law article into fundamental atomic propositions, as illustrated in Fig.~\ref{overallthinking}.

According to the proposed architecture, each law article is composed of triggers, judgements, and thresholds when digitized. 
Every formulation in traffic law is expressed using calculable formulas or logical comparisons, and the logical comparisons in triggers are defined. Then, for each comparison, a suitable threshold is set. Using the MTL, the law article is first decomposed into semantic elements and then each element is transferred into MTL expressions. After that, using all the definitions, the atomic propositions in MTL are further converted into triggers and logical judgements with thresholds. 

When digitizing traffic laws, it is necessary to express each law article’s trigger domain ranges using the MTL expressions and clarify the atomic propositions and their logical relationships used in the digitization process. Therefore, in a traffic scenario, the more scenes a specific law article involves, the more parameters or fuzzy parameters need to be determined based on the actual data in a digitized atomic proposition, the more complex the logical relationship between atomic propositions is, and the more difficult it is to digitize law articles. Based on different types of restrictions in traffic laws, this study subdivides 25 driving-related articles in \textit{The Regulation on the Implementation of the Law of the People's Republic of China on Road Traffic Safety} into more than 90 detailed clauses. The digitization difficulty of each clause is measured according to the scene fitness, the atomic proposition complexity, and the number of fuzzy thresholds. 

\begin{figure}[htbp]
    \centering
    \includegraphics[width=0.5\linewidth]{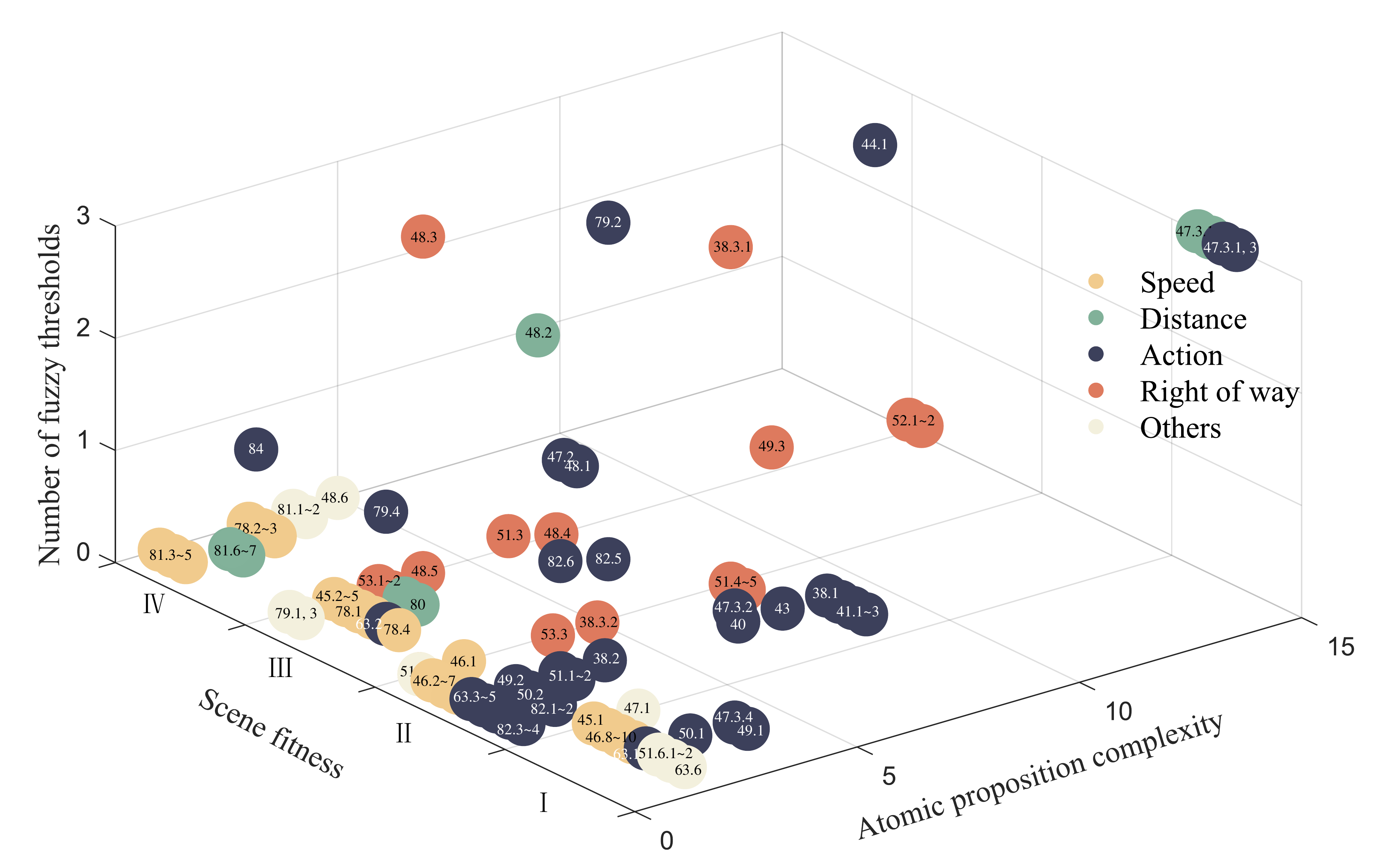}
    \caption{Digital complexity of regulations}
    \label{complexity}
\end{figure}

Among them, scene fitness is the scene level that describes how narrow a scene is when a particular clause takes effect. The scene fitness is divided into four levels denoted by I--IV. The higher the level is, the more restrictive conditions of the scene will be, and the more difficult in digitization at the triggering condition level. The atomic proposition complexity is the number of atomic propositions required for the logical judgement of clauses after the triggering scene is determined. The more atomic propositions are required, the more complex the logical judgement will be. At the same time, the more ambiguous the statutes describe the behavior, the more complex the logical judgement is required. As shown in Fig.~\ref{complexity}, all the driving relevant traffic law articles are evaluated, and the number in each circle represents the entry of the article, which can be found in our website in detail. 

The atomic propositions with a complexity of five or higher are basically motion and right-of-way restrictions because the speed and other clauses limitations can be indicated by the scene and thresholds and they do not involve interaction with other traffic participants. Action and right-of-way restrictions mainly need to consider the interactions with other traffic participants and include behaviors that a single atomic proposition cannot simply express, such as “keep the necessary safety distance”, and “slow down and drive to the right”. The more the interactions involved in a clause, the higher the atomic proposition complexity will be. Moreover, fuzzy thresholds are necessary for logical judgement in the digitization process, but these thresholds are not clearly defined by regulations. 
Therefore, how to define the fuzzy thresholds is a challenge. 
It is necessary to analyse reasonable thresholds from a large amount of data and explore their rationality. Thus, the number of fuzzy thresholds, the scence fitness and the atomic proposition complexity can affect digitization complexity to a different extent. As shown in Fig.~\ref{complexity}, the larger the index value is, the more complex the regulation’s digitization is from each aspect.

This paper takes the principal laws of highways and intersections as examples. All the logical definitions that the examples involved are listed in APPENDIX \ref{appendixc}, which are all calculable with given information at our best known.

\subsection{Highway and intersection traffic law digitization examples}
\subsubsection{Highway traffic law violation monitor}
This study selects the five most important and common traffic law articles related to the driving process on highways under non-adverse weather conditions from \textit{The Regulation on the Implementation of the Law of the People's Republic of China on Road Traffic Safety}. These five articles define the speed, distance and behavior constraints, and the original texts are as follows:

\noindent 1) Article-44 The motor vehicles changing lanes shall not affect the motor vehicles driving normally along the corresponding lanes.

\noindent 2) Article-47 After confirming there is a sufficient safe distance, the latter shall overtake the vehicle mentioned first from its left side, and after there is a second necessary safe distance between them, the overtaking vehicle shall turn on right turn light and return to the original lane.

\noindent 3) Article-78 Driving speeds for different vehicle lanes of an expressway shall be indicated, the maximum speed shall not exceed 120 kilometers per hour while the minimum speed shall not be lower than 60 kilometers per hour. Where there are two vehicle lanes in the same direction, the minimum speed for the left lane is 100 kilometers per hour; and where there are three vehicle lanes or more in the same direction, the minimum speed for the farthest left lane is 110 kilometers per hour, and 90 kilometers per hour for the middle lane. Where there is any discrepancy between the speed indicated by a speed limit sign put up on a road and the driving speeds mentioned above, a motor vehicle shall be driven at the speed indicated by the speed limit sign on the road.

\noindent 4) Article-80 Where a motor vehicle is running on expressway at a speed which exceeds 100 kilometers per hour, a distance of 100 meters or more shall be maintained from the vehicle in front in the same vehicle lane; and when the speed is lower than 100 kilometers per hour, the distance from the vehicle in front may be narrowed appropriately, but the minimum distance may not be less than 50 meters.

\noindent 5) Article-82.3 When driving a motor vehicle on expressway, the driver shall not drive over or on the dividing line of vehicle lanes or on the shoulder.
\begin{table}[htbp]
    \footnotesize
	\centering
	\caption{{\normalsize Trigger domain and logic judgements of typical Articles of highway}}
	\renewcommand{\arraystretch}{1.0}
	\label{table2}
	\begin{tabular}{p{3cm} p{4cm} p{6cm}}
		\toprule
		Article & Trigger domain & Logical requirement \\
		\midrule
		\multirow{1}{2cm} {Article44} & $\textit{T}_{\emph{ChangeLeftlane}}$ & \textit{ChangeLeftLaneCompliance} = $\bm T$\\
		& & $\Leftrightarrow$ \textit{\quad FrontViolation}=$\bm F$ \\
		& & \textit{$\qquad$ $\wedge$ RearLeftViolation}=$\bm F$ \\
		& & $\qquad$ $\wedge$ \textit{FrontLeftViolation} = $\bm F$\\
		& & $\qquad$ $\wedge$ {LngTmOnLine} = $\bm F$\\
		& $\textit{T}_{ChangeRightlane}$ &\textit{ChangeRightLaneCompliance} = $\bm T$\\
		& & $\Leftrightarrow$ \textit{\quad FrontViolation}=$\bm F$ \\
		& & \textit{\qquad $\wedge$ RearRightViolation}=$\bm F$ \\
		& & $\qquad \wedge$ \textit{FrontRightViolation} = $\bm F$\\
		& & $\qquad \wedge$ {LngTmOnLine} = $\bm F$\\
		\multirow{1}{2cm} {Article47} & $\textit{T}_{Overtake1}$ & \textit{FrontnotOvertake} = $\bm T$\\
		& & $\wedge$ \textit{ChangeLeftLaneCompliance} = $\bm T$\\
		& $\textit{T}_{Overtake2}$ & \textit{RecommendedSpeed} = $\bm T$\\
		& $\textit{T}_{Overtake3}$ & \textit{ChangeRightLaneCompliance} = $\bm T$ \\
		& $\textit{T}_{OvertakeR}$ & \textit{OvertakeonRight} = $\bm F$\\
		{Article78} & $\textit{T}_{SpeedLimit}$ & \textit{SpeedCompliance=} $\bm T$\\
		Article80 & $\textit{T}_{\emph{KeepFollowingDistance}}$ & \textit{FollowingCompliance}=$\bm T$\\
		{Article82.3} & $\textit{T}_{DriveonLaneline}$ & {LngTmOnLine}= $\bm F$ \\
		\bottomrule
	\end{tabular}
\end{table}

Article 82.3 ensures normal traffic flow, suggesting that vehicles shall not drive on the lane line for a long time. Article 44 is the law related to lane-changing behavior. Namely, when a vehicle changes driving lanes, the vehicle will drive on the line for a certain period. Therefore, lane-changing monitoring must include monitoring Article 82.3 to determine whether a vehicle is changing lanes too slowly. Article 47 is the law defining overtaking behavior. A complete overtaking behavior can be divided into three stages: the lane-changing stage, the target vehicle passing stage and the original lane-returning stage. The first and third stages include lane-changing behavior, so they also include monitoring Article~44. The traffic law monitor can be layered and modular based on those relationships above. The upper-layer modules can call the lower-layer modules to reduce the digitization work and computational power consumption.

The traffic law violation monitoring of each law article represents the combination and invocation of related modules, and each module includes the trigger domain and logical judgements. Table~\ref{table2} presents the trigger domain and logical judgements of laws considered in this study.

In order to make a better logical judgement, this study uses the combination of the global coordinate system and the ego vehicle coordinate system to facilitate the usage of information obtained from the HD map and a vehicle sensor. The lane to which \textit{obj} belongs is defined as Lane$({obj})$. The innermost lane of the road in the same direction is defined as Lane 1, and the IDs of other lanes and lane lines are increasing outward. The lane line with ID $i$ is expressed as LaneLine$(i)$, which is represented by a cubic fitting curve. In this study, the area around the ego vehicle is divided into six areas: front left area, rear left area, front area, rear area, front right area and rear right area. For vehicles in each area, only the vehicle nearest to the ego vehicle is considered. Use subscripts to express the position relationship of \textit{Tgt} relative to \textit{Ego}, for instance, $\emph{Tgt}_{f}$ represent \textit{Tgt} in the front area of \textit{Ego}. The subscripts of the other five areas are similar.

It should be noted that Article 78 should always be monitored when driving on the main way of the highway. Thus, the triggering condition that the ego vehicle is located on the main way of the highway is expressed as follows:
\begin{equation}
    \begin{aligned}
    T_{Speed Limit}\Leftrightarrow \text{RoadType}(Ego)=highway_{main}\\
    \end{aligned}
\end{equation}
where RoadType is the road type code (see Table S2), and highwaymain represents the main way of the highway.

The speed limit monitoring module is activated when the triggering condition is met, and it ensures that the vehicle's speed adheres to the traffic sign regulations or the lane-specific speed limit requirements.

\begin{equation}
    \begin{aligned}
    &\emph{SpeedCompliance} \Leftrightarrow\\
    & \left(\text{SpdSignArea}(Ego)\wedge vx(Ego)\in \left[V_{sign\_min}, V_{sign\_max}\right]\right)\\
    &\vee \left(\begin{array}{l}
    \neg \text{SpdSignArea}(Ego) \wedge vx(Ego)\in[60,120]\text{km/h}\\
    \wedge \left(\begin{array}{l}
    \left(\begin{array}
    {l}N_{mw}\geq3 \wedge \text{Lane}(Ego) \in \left[2,N_{mw}\right)\\
    \wedge vx(Ego)\geq90\text{km/h} 
    \end{array}\right)\\
    \vee \left(\begin{array}{l}
    N_{mw}\geq 3 \wedge \text{Lane}(Ego)=1 \\
    \wedge vx(Ego)\geq110\text{km/h}
    \end{array}\right)\\
    \vee \left(\begin{array}{l}
    N_{mw}=2 \wedge \text{Lane}(Ego)=1 \\
    \wedge vx(Ego)\geq100\text{km/h} 
    \end{array}\right)\\
    \end{array}\right)
    \end{array}\right)
    \end{aligned}
\end{equation}

where $V_{sign\_max}$ and $V_{sign\_min}$ represent the upper and lower limits indicated by the speed limit sign, respectively; $N_{mw}$ is the number of main ways in the same direction. 

For Article 80, the triggering condition is that the ego vehicle is driving on the main way of the highway, and there is a vehicle in front of it.
\begin{equation}
    \begin{aligned}
    T_{\emph{KeepFollowingDistance}}\Leftrightarrow &(\text{RoadType}(Ego)=highway_{main})
    \wedge \exists  \emph{Tgt}_{f}
    \end{aligned}
\end{equation}

When the triggering condition is satisfied, the following distance monitoring module starts monitoring. The corresponding mathematical expression is as follows:
\begin{equation}
    \begin{aligned}
    &\emph{FollowingCompliance} \Leftrightarrow \\
    &(vx(Ego)>100\text{km/h} \wedge \textit{distance}(Ego,\emph{Tgt}_{f})>100\text{m})\\
    &\vee (vx(Ego)\leq 100\text{km/h} \wedge \textit{distance}(\emph{Ego},\emph{Tgt}_{f})>50\text{m})
    \end{aligned}
\end{equation}
For Article 82.3, the triggering condition is that the ego vehicle overlaps with the lane line, which can be expressed as follows:
\begin{equation}
    \begin{aligned}
        &T_{DriveonLaneline}\Leftrightarrow \\
        & \text{overlap}({Area}(Ego), \text{LaneLine}(\text{Lane}({Ego})))\\
        & \vee \text{overlap}({Area}(Ego), \text{LaneLine}(\text{Lane}({Ego})+1))\\
    \end{aligned}
\end{equation}

When the triggering condition is satisfied, the driving-on-lane-line monitoring module starts monitoring. The moment when the triggering condition is satisfied is denoted by $t_{in}$. “Vehicle drives on the lane line for a long time” is defined as follows:
\begin{equation}
    \begin{aligned}
        &\text{LngTmOnLine}(Ego)
        \Leftrightarrow P_{[t_{in},t_{now}]}(T_{DriveonLaneline}) \wedge t_{now}-t_{in}>t_{max\_cl}
    \end{aligned}
\end{equation}
where $t_{now}$ is the current time and $t_{max\_cl}$ is the maximum allowable time for driving on the lane line.

For Article 44, taking changing lanes to the left as an example. Trigger conditions are defined as the point when the ego vehicle decides to change lanes to the left and overlaps with the left lane line. During this maneuver, the ego vehicle must maintain a safe distance from other relevant vehicles. These include the front vehicle, left-front vehicle and left-rear vehicle. As mentioned earlier, the lane-changing process involves driving along the lane line for a certain period. Therefore, the lane-changing monitoring module must include monitoring of driving on the lane line, ensuring that the vehicle does not remain on the lane line for an extended duration during the lane-changing process. The corresponding mathematical expression is as follows:
\begin{equation}
    \begin{aligned}
    &\textit{T}_{\emph{ChangeLeftlane}} \Leftrightarrow \\
    & \text{Decision}({Ego}) = \emph{ChangeLeftlane}\wedge vy(Ego)>0\\
    & \wedge \text{overlap}({Area}(Ego), \text{LaneLine}(\text{Lane}({Ego})))\\ 
    \end{aligned}
\end{equation}

When the triggering condition is satisfied, the changing-left-lane monitoring module starts monitoring.
As defined by Article 44 and presented in Table~\ref{table2}, longtime\_on\_laneline belongs to the driving-on-lane-line monitoring module, and it will be called directly when the changing-left-lane monitoring module starts to monitor. The specific definitions of the other three sub-propositions are as follows:

\begin{equation}
    \begin{aligned}
    &\emph{FrontViolation} \Leftrightarrow \\
    & \exists \emph{Tgt}_{f}\wedge\left(\begin{array}{l}
    {TTCX}(Ego, \emph{Tgt}_{f})\leq TTCx\\
    \vee \ \textit{distance}(Ego, \emph{Tgt}_{f})\leq d_{clmin}
    \end{array}\right)\\
    \end{aligned}
\end{equation}
\begin{equation}
    \begin{aligned}
    &\emph{RearLeftViolation} \Leftrightarrow\\
    & \exists \emph{Tgt}_{rl}\wedge\left(\begin{array}{l}
    {TTCX}(\emph{Tgt}_{rl}, Ego)\leq TTCx\\
    \vee \ \textit{distance}(\emph{Tgt}_{rl}, Ego)\leq d_{clmin}
    \end{array}\right)\\
    \end{aligned}
\end{equation}
\begin{equation}
    \begin{aligned}
    &\emph{FrontLeftViolation} \Leftrightarrow\\
    & \exists \emph{Tgt}_{fl}\wedge\left(\begin{array}{l}
    {TTCX}(Ego, \emph{Tgt}_{fl})\leq TTCx\\
    \vee \ \textit{distance}(Ego, \emph{Tgt}_{fl})\leq d_{clmin}
    \end{array}\right)\\
    \end{aligned}
\end{equation}

In this study, the minimum actual distance and time to the collision are used to define the judgement threshold of compliance distance; $TTCx$ is the minimum time to longitudinal collision; $d_{clmin}$ is the minimum actual distance of a vehicle from relevant vehicles. In addition, it is needed to monitor $\text{LngTmOnLine}$.

Further, according to Article 47, the overtaking process is divided into three consecutive stages and each of them is monitored. Namely, overtake stage 1 is the lane-changing stage, where the ego vehicle can change lanes to the left when overtaking. Thus, the monitoring module of overtake stage 1 should include the monitoring module of changing the left lane. Overtake stage 2 is the target vehicle passing stage. In this stage, it is recommended to maintain a proper speed difference between the ego vehicle and the target vehicle. Overtake stage 3 is the original lane returning stage, where the ego vehicle changes lanes to the right to return to its original lane. Therefore, the monitoring module of overtake stage 3 should also include the monitoring module of changing the right lane. It should be noted that overtaking from the right is not allowed. The detailed violation monitor of Article 47 is given in APPENDIX \ref{appendixa}.

\subsubsection*{Intersection traffic law violation monitor}
At intersections, traffic laws are mainly defined by traffic lights and the right of way. In this paper, the traffic laws at intersections are categorized into three groups: traffic-light rules, virtual-lane-follow rules, and right-of-way rules, which are defined by Articles 38--41. Accordingly, Article 38 is divided into six sub-articles, which are shown in Table~\ref{table3}.

\begin{table}[htbp]
  \footnotesize
  \centering
  \caption{The sub-articles of Article 38}
  \renewcommand{\arraystretch}{1.0}
  \begin{tabular}{p{3cm} p{8cm} }
  \toprule
  \multicolumn{1}{c}{Article 38}  & \multicolumn{1}{c}{Content of regulation} \\
  \midrule
    \multirow{3}{2cm} {Traffic lights rules}  & Green light means that vehicles are allowed to proceed; \\
     & Yellow light means that vehicles across the stop line may keep on driving; \\
     & Red light means that vehicles are prohibited from passing \\ \hline
    \multirow{1}{2cm} {Virtual lane follow rules}  & Vehicles shall try to follow the best theoretical route when passing through intersections \\ \hline
    \multirow{4}{2cm} {Right of way rules}  & The making-a-turn vehicles shall not interfere with the straight-moving vehicles and pedestrians that are allowed to pass; \\
     & At the red light, the right-turn vehicles may proceed without interfering with other vehicles and pedestrians that are allowed to pass \\
  \bottomrule
  \end{tabular}
  \label{table3}
 \end{table}

\begin{figure}[htbp]
    \centering
    \includegraphics[width=0.5\linewidth]{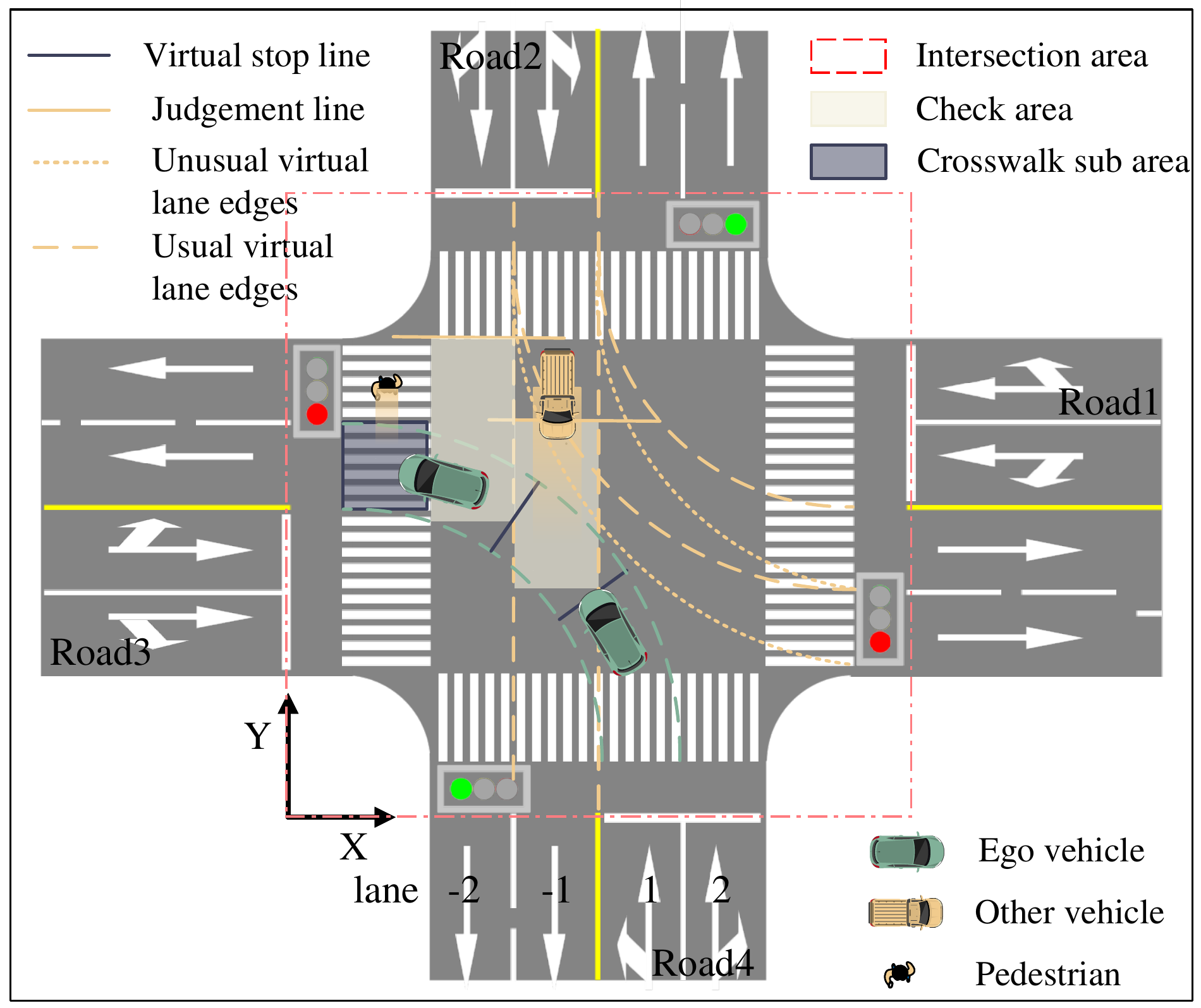}
    \caption{Map information of intersection}
    \label{cross}
\end{figure}

Among the traffic lights rules, only red and yellow lights restrict the passage of vehicles. During the red light phase, vehicles are prohibited from entering the intersection. During the yellow light phase, vehicles that have not yet entered the intersection are not permitted to do so. According to the virtual-lane-follow rules, vehicles should drive inside the virtual lane, which is calculated by an algorithm based on the map information. Virtual lane defines which areas a vehicle can drive in and which areas are not recommended for driving. If a vehicle does not drive within these areas, the vehicle will break the law. The right-of-way rules are the most complex rules among all rules related to intersections. They define that a vehicle should avoid high right-of-way vehicles, pedestrians, and non-motor vehicles. To monitor compliance of the right-of-way rules, virtual stop lines and check lines are defined to determine if a target vehicle is a high right-of-way vehicle and where the ego vehicle should stop. The map information required for compliance monitoring is presented in Fig.~\ref{cross}.

In Fig.~\ref{cross}, the intersection area is defined as the dashed-line rectangular area enclosed by the stop line, and the virtual lanes include the Usual and Unusual virtual lanes, representing the optimal and compliant but not recommended paths through the intersection, respectively. For every entrance lane, there is the corresponding direction-judgement line that is used to determine the intention of moving direction for each passing vehicle. When the ego vehicle makes a moving direction decision, the virtual lane is then determined, and the other virtual lane with a higher right-of-way that intersects with this virtual lane will be selected, the corresponding virtual stop line and check area will be generated. If there exists a target vehicle with a higher right-of-way passing through the check area, the ego vehicle should not pass the virtual stop line to ensure unobstructed passing of the target vehicle. Based on the direction-judgement line, when it is determined that another vehicle intends to drive along this lane, and the midpoint of the front end of this vehicle is within this check area, it is considered that there is a high right-of-way vehicle in the check area, and the ego vehicle shall not pass the virtual stop line. The crosswalk sub-area is part of the crosswalk corresponding to the lane width. Its purpose is to determine whether passing vehicles impede nearby pedestrians and non-motor vehicles. The virtual elements used for monitoring and judgement are shown in Fig.~\ref{cross}.The detailed violation monitor of Article 38 is listed in APPENDIX~\ref{appendixb}.

\section{Monitor validation and data analysis}
\subsection{Vehicle trajectory datasets and high precision map}
All aforementioned MTL expressions of highway and intersection traffic laws were programmed in MATLAB/Simulink using different enable subsystems as part of the proposed online monitor. The proposed traffic law violation monitor was verified on Chinese vehicle trajectory datasets.

The highway monitor was verified on the AD4CHE dataset~\cite{zhang2021aerial}.
It is worth mentioning that the AD4CHE dataset includes congested sections of highways, so in this dataset, vehicles drive at a relatively slow speed, which results in high traffic law violation rates of vehicle speed and distance.
The intersection monitor was verified on the SIND dataset~\cite{xu2022drone}. The SIND dataset was collected at a signalised intersection in China, and it contains information on traffic participant trajectories, traffic light status and high-definition maps.

As the proposed monitor is a vehicle-side online monitor, the data from the highway monitor test were first transferred into the ego vehicle's view. Once the ego vehicle is selected, the vehicle coordinate system is established according to the ego vehicle's states, and all the other vehicles' coordinates are transferred into this coordinate system. Considering that not all vehicles could provide complex information about themselves and the surrounding environment, this study aimed to use a minimal amount of data to make the monitor run to improve the applicability of the proposed monitor. The ego vehicle’s parameters (e.g., width and length) and states (e.g., velocities and accelerations in the \textit{X} and \textit{Y} directions and head angle) were combined as the ego vehicle's data bus. The other vehicles’ data (e.g., relative coordinates, velocities, accelerations, and dimensions) that could be easily obtained by mature algorithms were combined and used as an object data bus. In addition, each point of the road lines was transferred and used in a map data bus. Using these data buses as inputs, the highway monitor could monitor the law violation behaviors of any vehicle from the dataset.

Meanwhile, in the intersection monitor test, the data maintained the global coordinate system as the high-definition map was used. The bus also includes the status of pedestrians (e.g., relative coordinates and velocities) and Traffic light status. In addition, an extra intersection map data bus was constructed for the intersection monitor test. The intersection bus included roads, signs, stop lines and crosswalk. The road data provided information on the lane ID, lane mark and lane direction (i.e., enter or leave the intersection). For each lane, there were right and left line types, points, and line ending point coordinates used to describe lanes in detail. There were also line ID, types, start and ending points coordinates for the stop lines. And for the crosswalk areas, there were IDs and vertex coordinates. Using the constructed data buses, the monitors could effectively monitor traffic law violation behaviors of the ego vehicle.

It should be mentioned that there was no decision information on any vehicle in the dataset. Therefore, for each vehicle selected as an ego vehicle, the behavior decision was approximated according to the states and global trajectory of the vehicle. All decisions of selected vehicles were generated using the following methods.

For the highway traffic law violation monitor, it was challenging for the online monitor to distinguish whether a vehicle would perform a lane change or an overtake based on vehicle states only. Nevertheless, both a lane change and an overtake involve lane-changing behavior. Consequently, the initial consideration was that the ego vehicle made a lane-changing decision when its lateral velocity exceeded 0.25 m/s.\cite{konigshof2022parameter}.
\begin{equation}
    \begin{aligned}
    &{vy(Ego)}>0.25m/s\Rightarrow \text{Decision}({Ego}) = \emph{ChangeLeftlane}
    \end{aligned}
\end{equation}
\begin{equation}
    \begin{aligned}
    &{vy(Ego)}<-0.25m/s\Rightarrow \text{Decision}({Ego}) = ChangeRightlane
    \end{aligned}
\end{equation}

Further, when there was a vehicle in front of the ego vehicle and their TTC was less than 20 s, and the lateral speed of the ego vehicle was higher than 0.25 m/s, we consider the decision made by the ego vehicle to be an overtaking rather than a lane change maneuver.
\begin{equation}
    \begin{aligned}
    &\left(\begin{array}{l}
    \exists \emph{Tgt} \in \text{front}(Ego) \wedge vx(\emph{Tgt})<vx(Ego) \\
    \wedge {TTCX}(Ego, \emph{Tgt})<20s \\
    \wedge|vy(Ego)|>0.25m/s
    \end{array}\right) \Rightarrow \text{Decision}({Ego}) = Overtake\\
    \end{aligned}
\end{equation}

For the intersection traffic law violation monitor, the decision information that needed to be obtained was the driving direction decision. In SIND dataset, the vehicle’s road ID in the first and last frames showed how the vehicle went. For the ego vehicle, the road ID of the first frame was denoted by $\text{RoadID}_{in}(Ego)$, and the road ID of the last frame was denoted by $\text{RoadID}_{out}(Ego)$.  According to the relationship between these two IDs, the direction decision of a vehicle was determined as follows:

\noindent 1. Go-straight decision:
\begin{equation}
    \begin{aligned}
    &\text{RoadID}_{out}(Ego)=(\text{RoadID}_{in}(Ego)+2)\|(\text{RoadID}_{in}(Ego)-2)\Rightarrow \text{Decision}({Ego}) = \emph{GoStraight}\\
    \end{aligned}
\end{equation}

\noindent 2. Turn-left decision:
\begin{equation}
    \begin{aligned}
    &\text{RoadID}_{out}(Ego)=(\text{RoadID}_{in}(Ego)+3)\|(\text{RoadID}_{in}(Ego)-1)\Rightarrow \text{Decision}({Ego}) = \emph{TurnLeft}\\
    \end{aligned}
\end{equation}

\noindent 3. Turn-right decision:
\begin{equation}
    \begin{aligned}
    &\text{RoadID}_{out}(Ego)=(\text{RoadID}_{in}(Ego)+1)\|(\text{RoadID}_{in}(Ego)-3)\Rightarrow \text{Decision}({Ego}) = \emph{TurnRight}\\
    \end{aligned}
\end{equation}

\subsection{Compliance thresholds and statistical analysis of results}

\subsubsection{Compliance thresholds}
In the traffic law digitization process, setting a proper threshold represents an essential part. A suitable threshold will make the monitor more reasonable and accurate. However, some thresholds are not defined by traffic laws, so how to select appropriate values has been the main challenge. In this study, the driving habits of human drivers in the dataset were analysed in the law article relative to driving scenarios to obtain the compliance behavior threshold with a certain degree of confidence.

In this paper, for highway traffic law violation monitoring, three thresholds were used. However, there have been no clear definitions for thresholds in the law articles, such as the minimum distance ($d_{clmin}$) to other vehicles when changing the lane, the maximum allowed time of driving on the line ($t_{max\_cl}$) and the minimum time to longitudinal collision ($TTCx$) with other vehicles when changing the lane. Using the AD4CHE dataset, the traffic law concerning vehicle driving behaviors was analysed, and the concerned thresholds were calculated. 

\begin{figure}[t!]
    \centering
    \subfigure[distance-deceleration scatter diagram]{
    \label{dmin}
    \includegraphics[width=5.8cm]{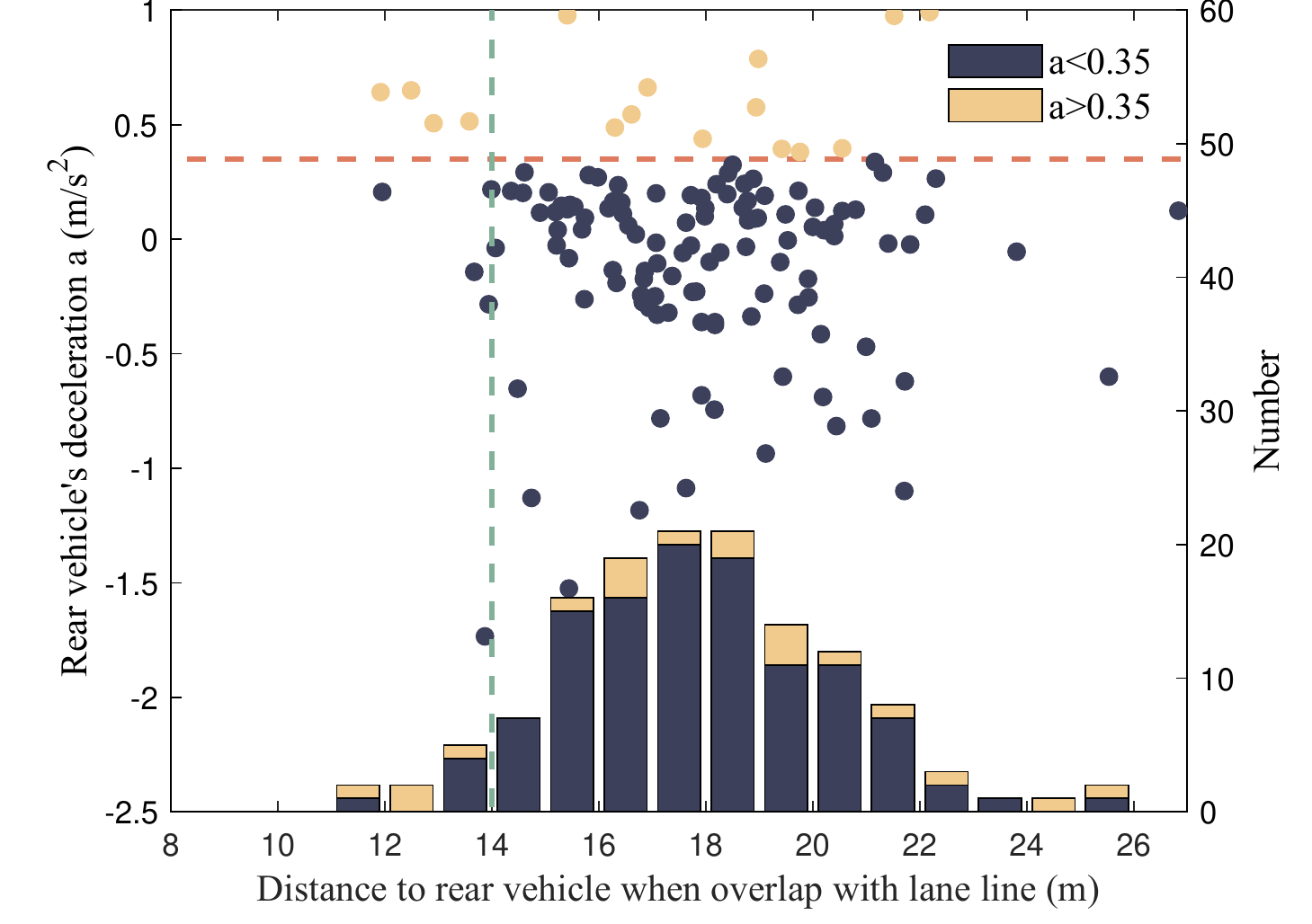}}
    \subfigure[beta fitting histogram of $T_{cl}$]{
    \label{Vycl}
    \includegraphics[width=5cm]{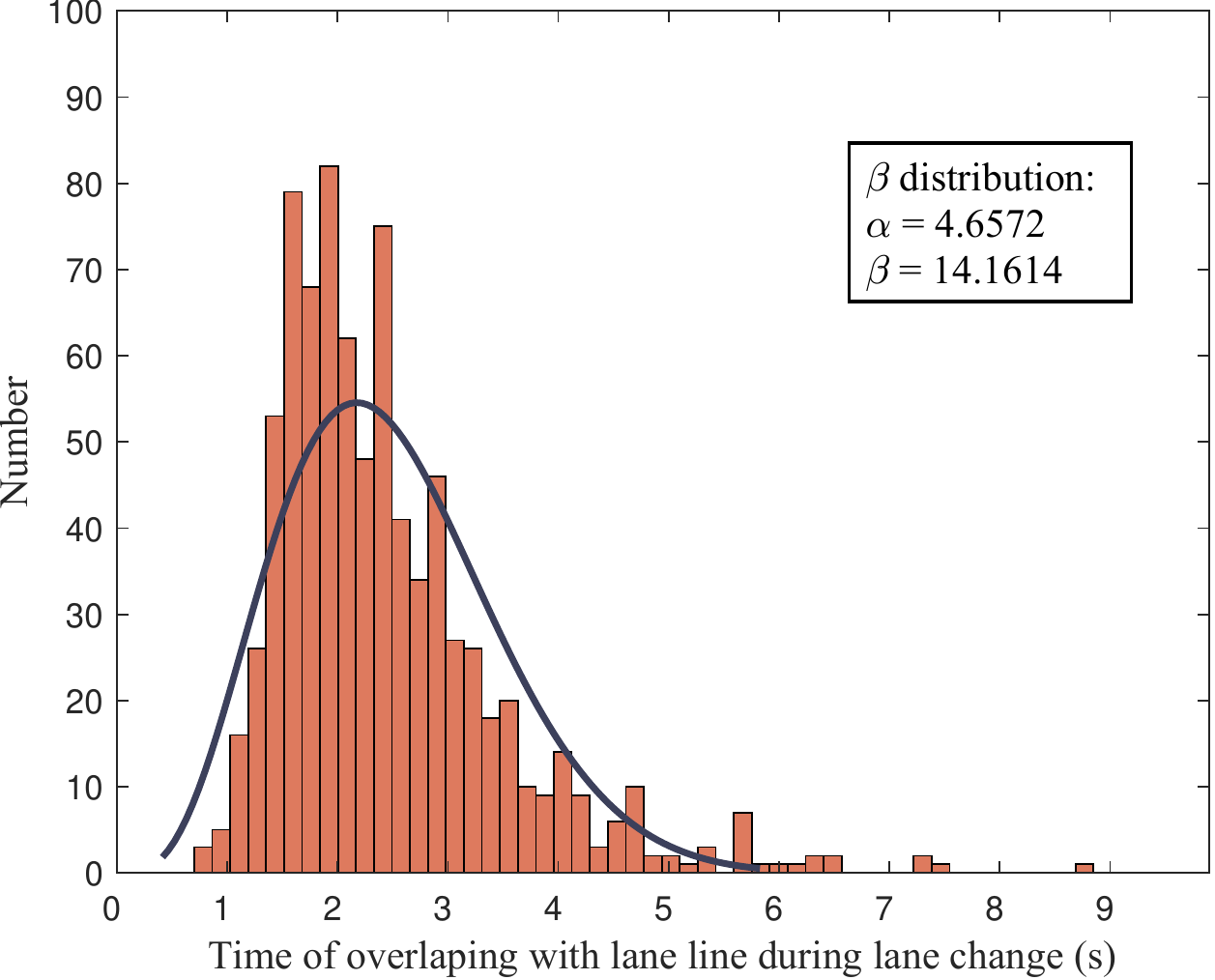}}
    \subfigure[TTC-$T_{cl}$/TTC scatter diagram]{
    \label{ttcx}
    \includegraphics[width=5cm]{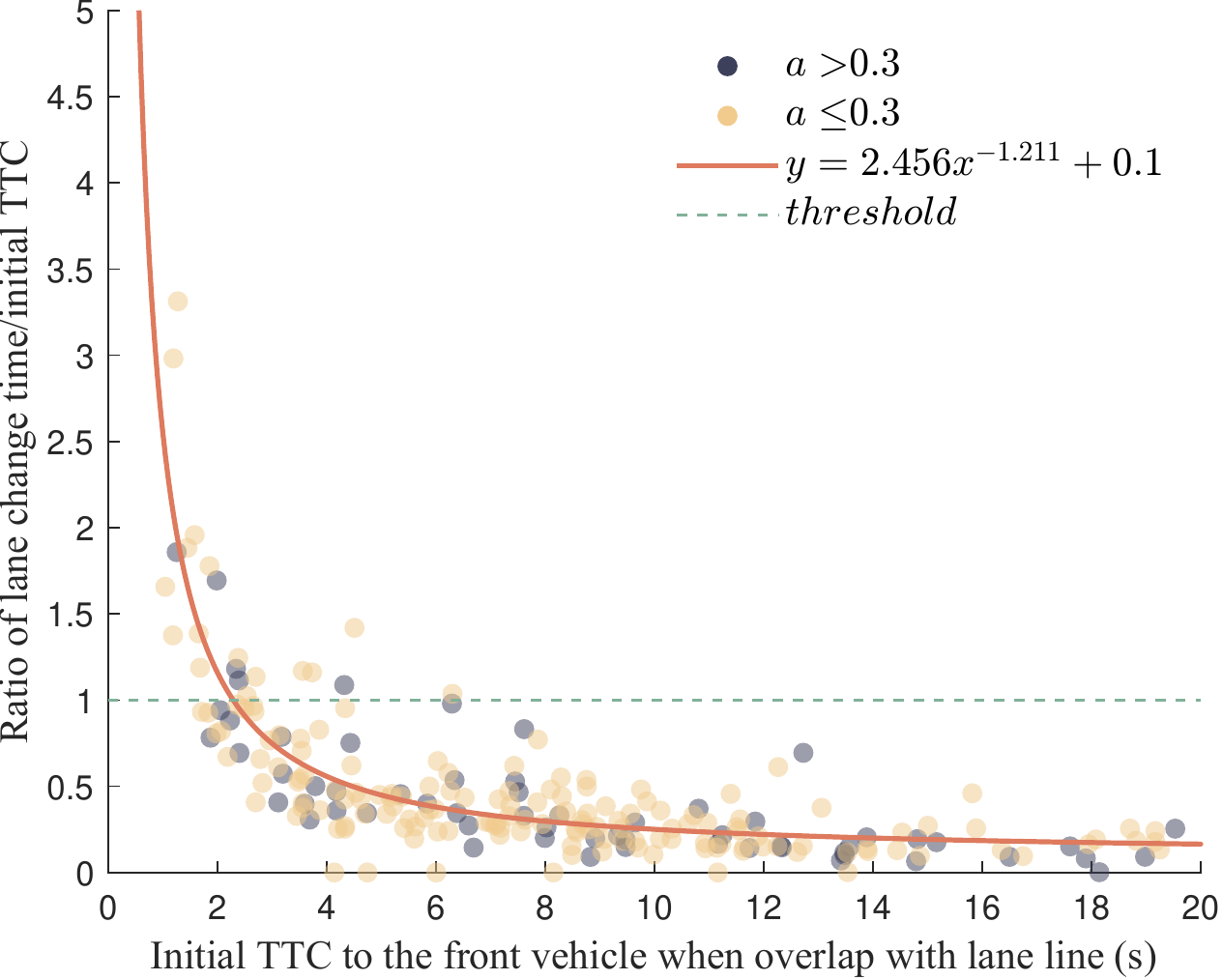}}
    \caption{Statistical results of threshold analysis}
    \label{threshold}
\end{figure}

Considering that an acceptable lane change progress should not affect other vehicles’ driving behaviors, the distances from behind vehicles and their deceleration behaviors can show whether they are affected by the front vehicle. To calculate the $d_{clmin}$ value, all effective lane-changing behaviors in the dataset were used. When the lane change began, the data on a particular scenario were extracted if a vehicle in the target lane appeared within 30 m behind. A scatter plot figure is presented in Fig.~\ref{dmin}, where the horizontal axis indicates the distance from the vehicle behind at the beginning of the lane-changing process, and the vertical axis represented the average deceleration of vehicles behind the target lane during the ego vehicle's lane changing. As shown in Fig.~\ref{dmin}, the distances between the ego vehicle and the vehicles behind were between 14~m and 20~m when the ego vehicle started to change lanes and overlapped with the lane line. Namely, most drivers decided to enter the target lane when they were 14~m--20~m away from the vehicle behind. The traffic law required that when another vehicle cut in, the ego vehicle needed to slow down properly and keep to the right to avoid the collision if conditions permit. Thus, the driver would decelerate properly for careful driving when a vehicle was cut in. As shown in Fig.~\ref{dmin}, at the speed of 0.35 m/s$^2$, there was an obvious deceleration dividing line. Namely, the deceleration values of most vehicles behind the ego vehicle were below 0.35 m/s$^2$ when the other vehicle cut in. The statistical graphic of vehicles' distances is presented at the bottom of Fig.~\ref{dmin}, where it can be seen that within the distance range of~14 m--20~m, the number of behind vehicles that decelerated less than 0.35~m/s$^2$ was larger than the number of behind vehicles that decelerate more than 0.35~m/s$^2$. Therefore, it was considered that when the dclmin was 14 m, and the other thresholds were met, the behind vehicles would not be obstructed by the lane-changing vehicles but only decelerate for prudent driving.

Furthermore, the time of vehicle overlapped with the lane line during the complete lane-changing behavior was calculated, and the statistical results are shown in Fig.~\ref{Vycl}. The vehicle-line overlapping time during the lane change varied from 0.8~s to 8.8~s, but most of them were concentrated within the range of 1.5~s--3~s, which obeyed the Beta distribution. A threshold of 6~s was selected as $t_{max\_cl}$ because 99.92\% of all vehicles’ lane changings could be completed within that time. Moreover, considering that the lane-changing time should be longer than the TTC to the front vehicle when changing the line, the initial TTC to the front vehicle and the ratio between the lane change time and the initial TTC were calculated, as shown in Fig.~\ref{ttcx}. Considering that the TTC to the front vehicle should be at least longer than the lane changing time, a normal ratio threshold of one was selected, and the fitting curve of the ratio was drawn. The intersection point indicated that the $TTCx$ should be 2.3~s.

\subsubsection{Statistical results}
\begin{figure}[htbp!]
    \centering
    \subfigure[Statistical results of the illegal cases]{
    \label{highwaynumber}
    \includegraphics[width=0.5\linewidth]{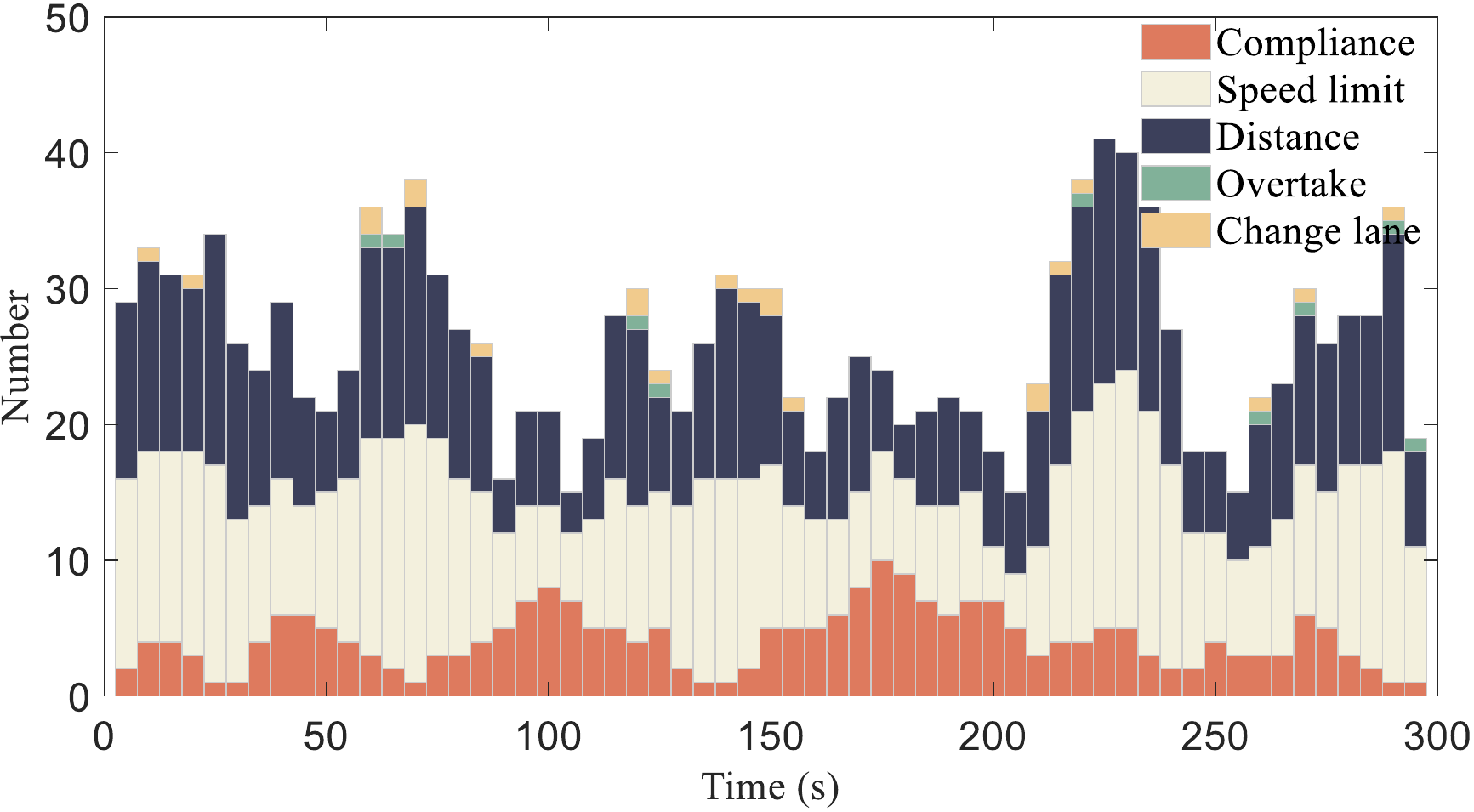}}
    \subfigure[Statistical results of the proportion of violation types]{
    \label{highwaypercentage}
    \includegraphics[width=0.4\linewidth]{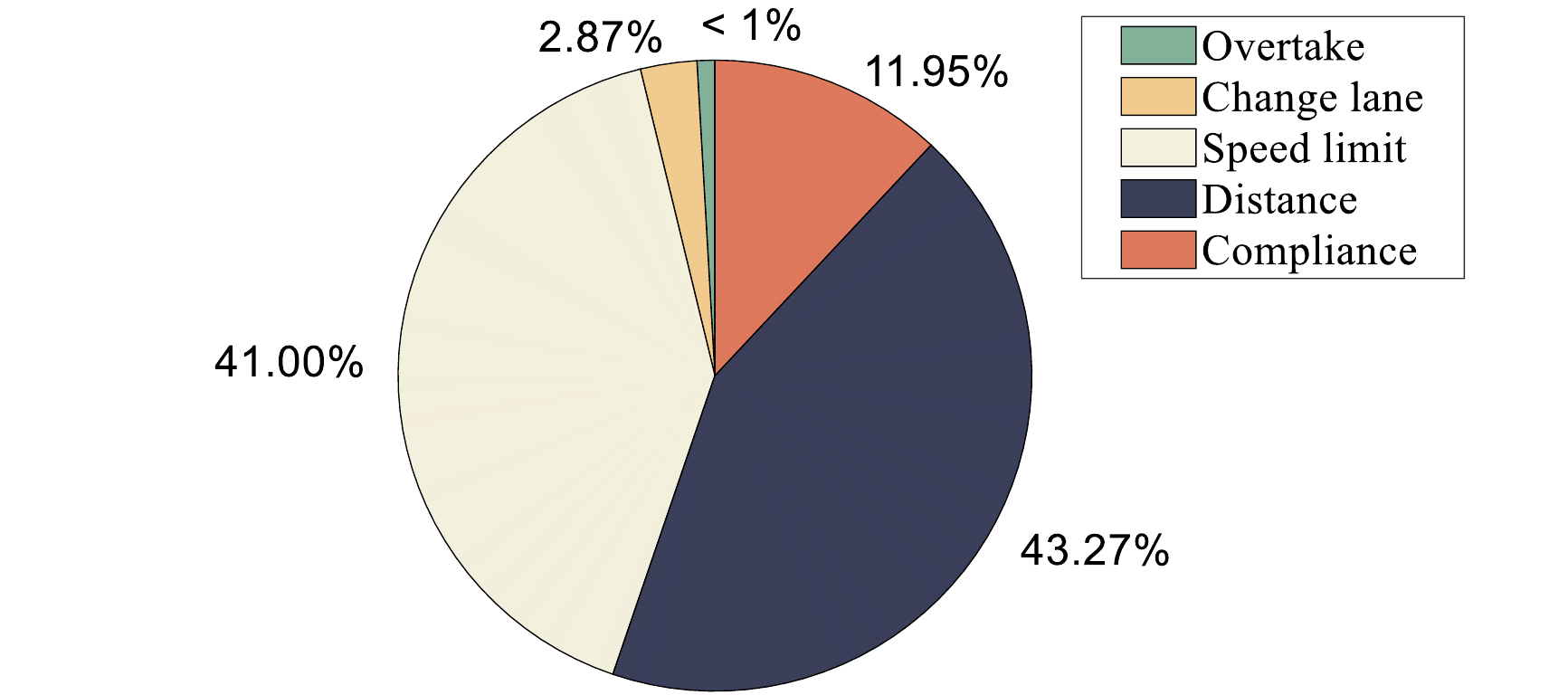}}
    \caption{Statistical results on the AD4CHE dataset fragment.}
    \label{results_hw}
\end{figure}

According to the set thresholds, traffic law violation monitoring was performed on each vehicle in the AD4CHE dataset. All vehicles in the dataset were marked by an ID. The IDs were assigned in a sequence; the selected vehicle was treated as the ego vehicle, while other vehicles in the scenario were treated as surrounding vehicles. The traffic law violation monitor monitored the ego vehicle’s behavior. After the last vehicle’s behavior was monitored, the statistical results of different traffic law violation types on the dataset were obtained. The violation statistical results of the 25th fragment in the AD4CHE dataset are presented in Fig.~\ref{results_hw}. This fragment lasted for 290 s; since the right part of the dataset was severely congested while the left part was better, only vehicles in the left part were monitored. The statistics of various types of highway traffic law violations on every 5 s are presented in Fig.~\ref{highwaynumber}. The proportion of each violation in the dataset is displayed in Fig.~\ref{highwaypercentage}. Due to the impact of traffic congestion, it was difficult for vehicles to reach the minimum speed limit specified on the highway. Maintaining compliance following distance was also challenging. Therefore, there were many violations of the speed limit and following distance. In contrast, there were fewer violations of overtaking and lane changing.

\begin{figure}[htbp!]
    \centering
    \subfigure[Statistical results of the illegal cases]{
    \label{crossnumber}
    \includegraphics[width=0.5\linewidth]{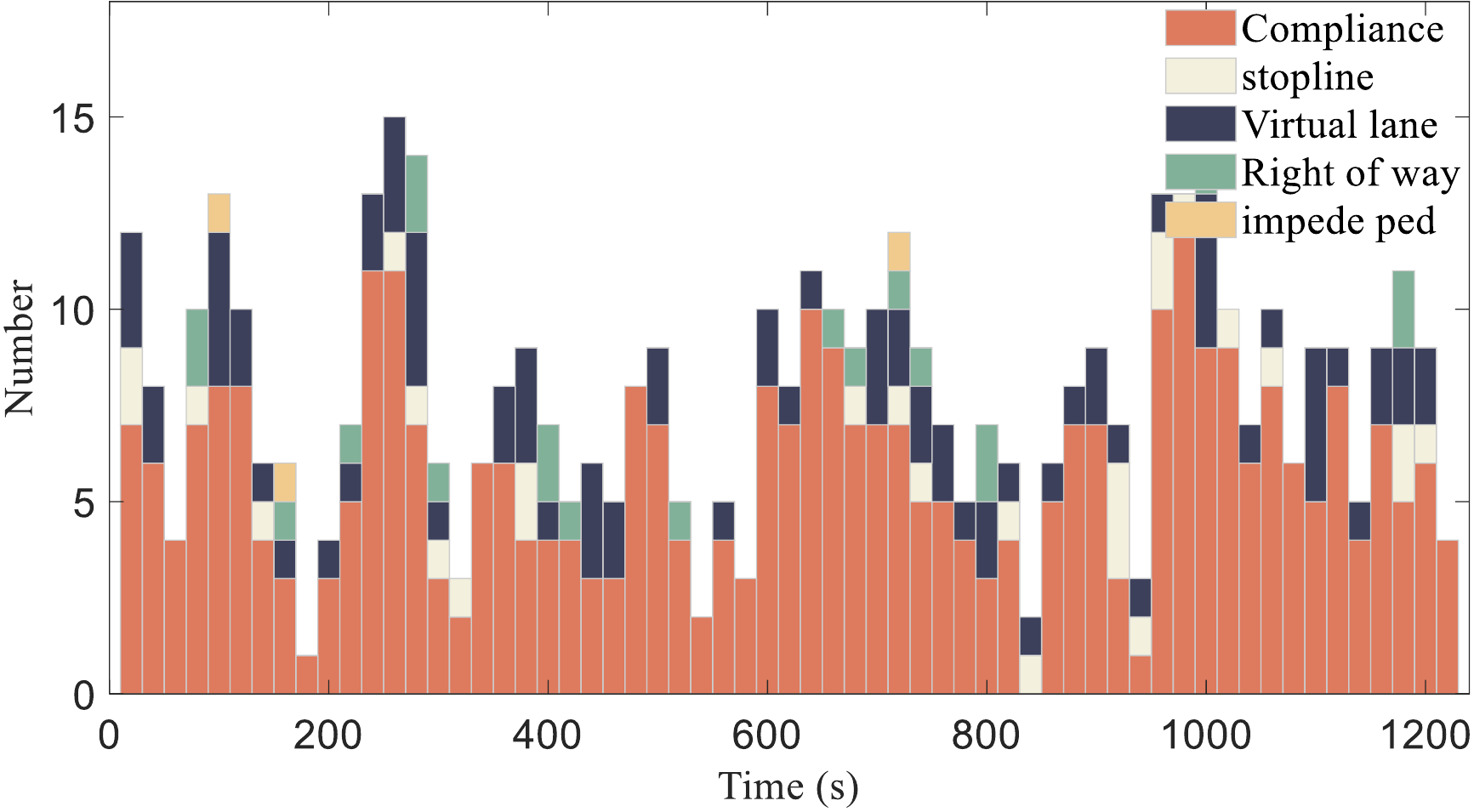}
    }
    \subfigure[Statistical results of the proportions of different violation types]{
    \label{crosspercentage}
    \includegraphics[width=0.4\linewidth]{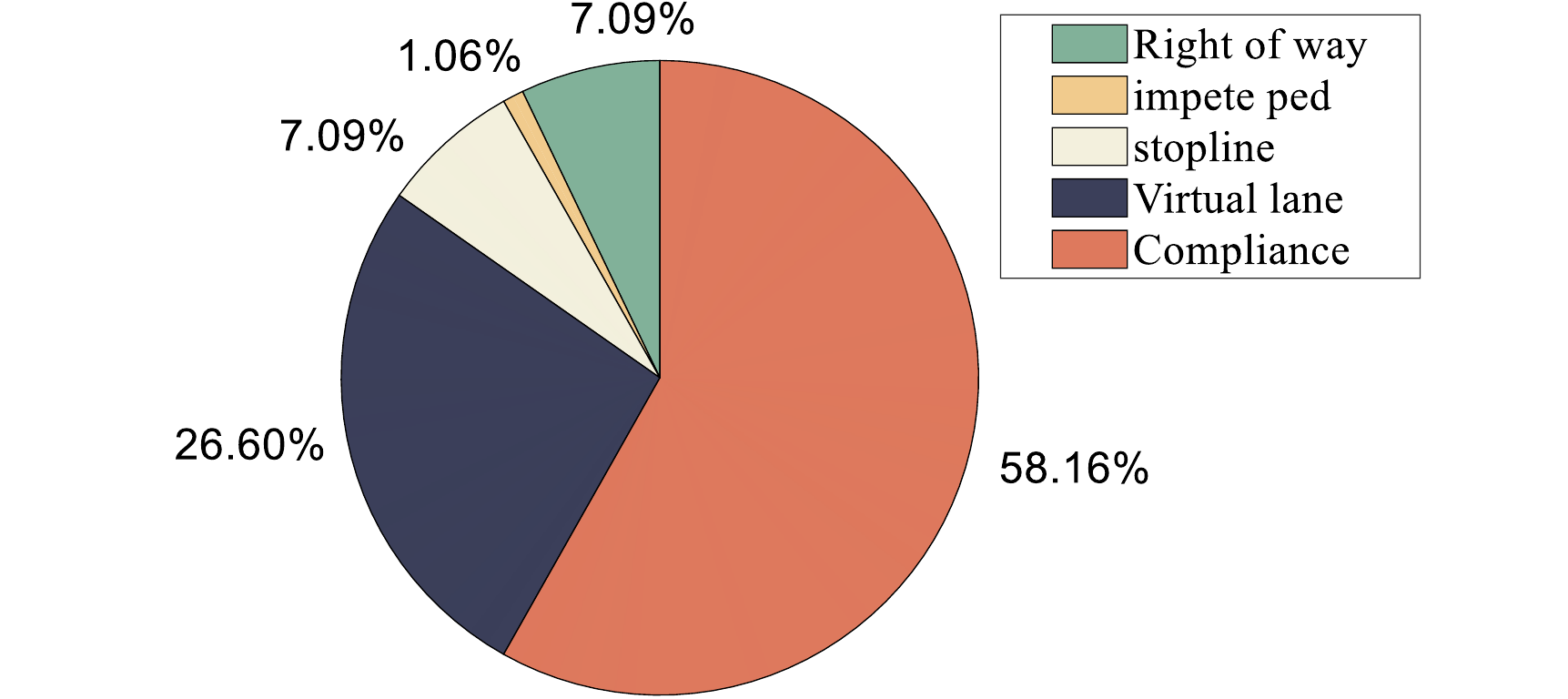}} 
    \caption{Statistical results on the SIND dataset fragment.}
    \label{results_c}
\end{figure}

No ambiguous threshold was introduced in the digitization process of intersection-related traffic laws. Similar to the monitoring test on the AD4CHE dataset, every vehicle in the SIND dataset was monitored for traffic law violations. The statistical results of the 8\_2\_1 fragment in the SIND dataset are presented in Fig.~\ref{results_c}; this fragment lasted for approximately 1,200~s. The statistics of intersection violations for each type on every 20~s are presented in Fig.~\ref{crossnumber}. Fig~\ref{crosspercentage} shows the proportions of different violations in the entire dataset. Since the virtual lanes were set as a circular arc without any relaxation value, the violation of this law was relatively high, and the follow-up research could provide a better virtual lane range. In addition, China has imposed severe punishment on the red light running while the yellow light running is treated more leniently. However, all traffic violations in the statistical result were yellow light running. The statistical results also showed that Chinese drivers' compliance with traffic laws was directly proportional to safety and punishment for violations.

\subsubsection{Illegal examples}
\begin{figure}[htbp!]
    \centering
    \includegraphics[width=0.6\linewidth]{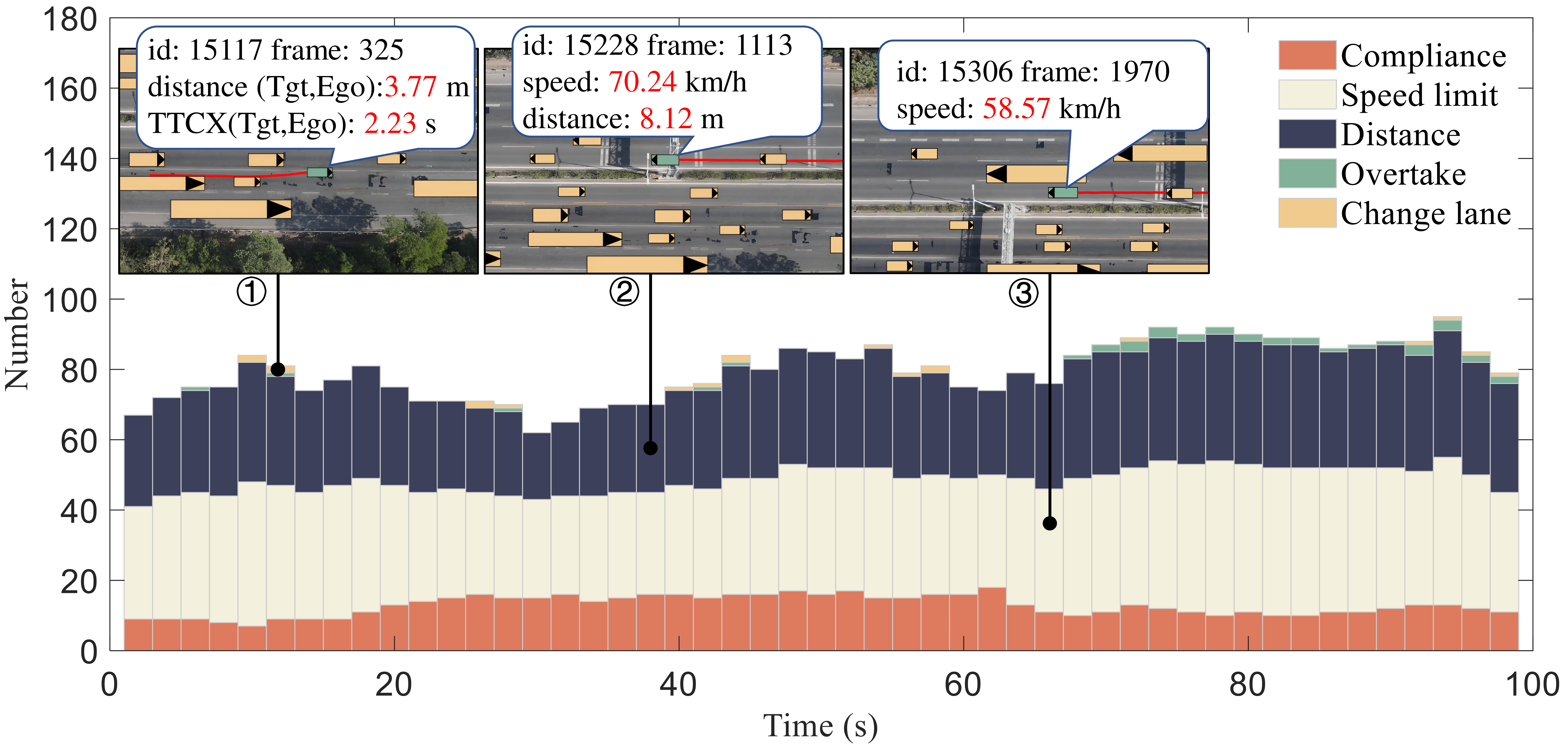}
    \caption{Illegal examples of highway}
    \label{example}
\end{figure}

The scenario of each dataset was reproduced by the visualisation program. Through the corresponding violation fragments in the visualisation program, it was more intuitive to verify the monitoring program's accuracy. Scenarios of vehicles violating the traffic law in the 44th fragment of the AD4CHE dataset are presented in Fig.~\ref{example}. Among them, the ID 15117 vehicle violated the lane-changing law due to its low TTCx value and distance from the vehicle behind. The ID 15228 vehicle ran at a speed of 70.24~km/h, and its following distance was only 8.12~m, which was much less than required. The ID 15306 vehicle ran at the innermost lane at a speed of 58~km/h, which was much lower than the minimum speed limit of 110~km/h.

\begin{figure}[htbp!]
    \centering
    \includegraphics[width=0.6\linewidth]{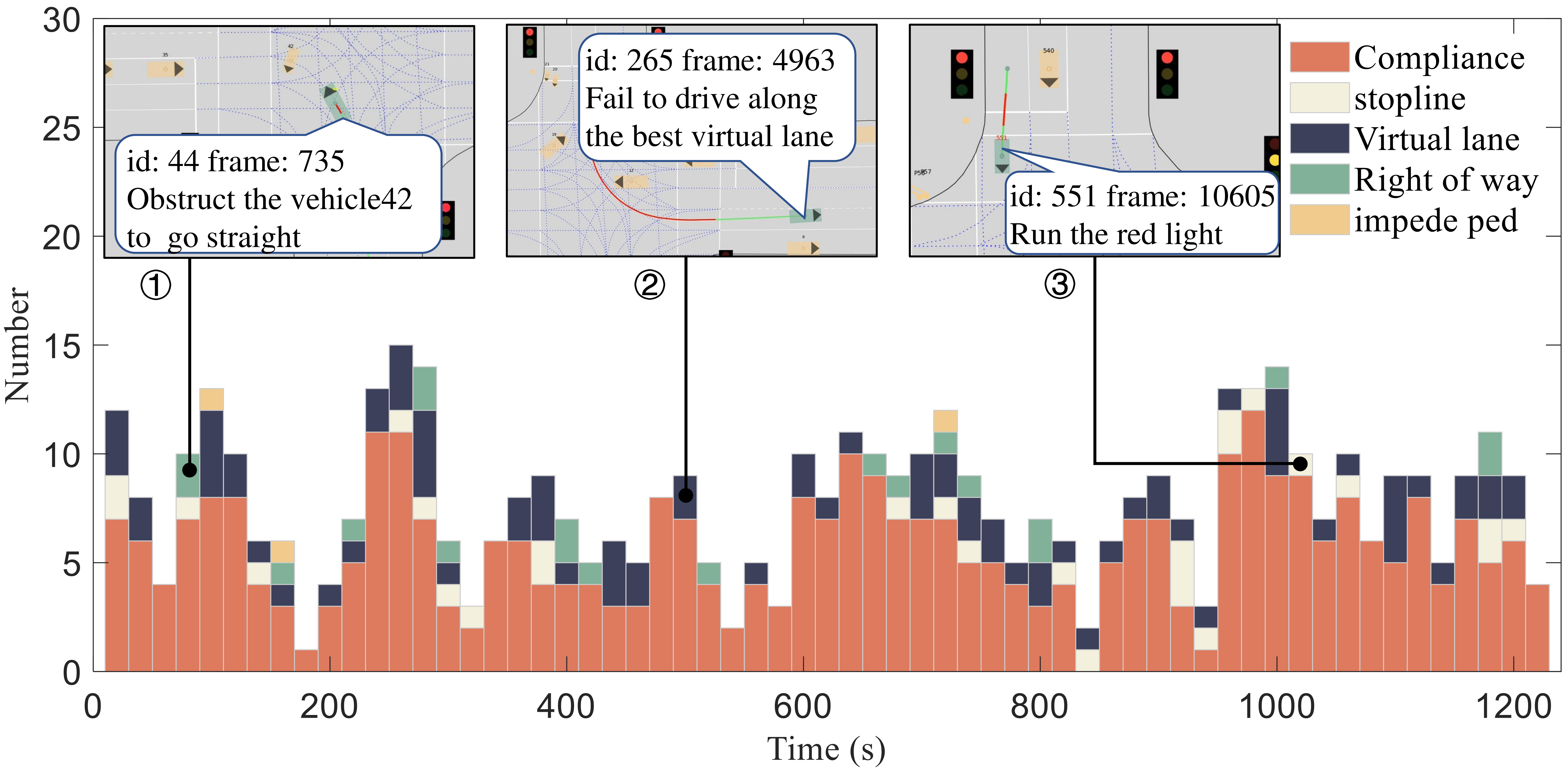}
    \caption{Illegal examples of intersection}
    \label{example2}
\end{figure}

Further, Fig.~\ref{example2} displays a number of scenarios where vehicles violated the traffic law in the fragment 8\_2\_1 in the SIND dataset. Among them, the ID 44 vehicle violated the right of way; it obstructed the straight ahead of the ID 42 vehicle. The ID 218 vehicle violated the virtual lane rule. Namely, when passing the intersection, this vehicle did not follow the best virtual lane. Even in some periods of time, it did not follow the not-recommended virtual lane. The ID 551 vehicle violated the traffic light rules. It crossed the stop line at the red light.

\section{Conclusion and future work}
This paper proposes an online traffic law violation monitor for monitoring an AV’s behaviors. The proposed online monitor runs with the autonomous system and provides the traffic law violation results in real-time based on the information from the vehicle’s perception and decision systems. The obtained results can be further used for making better law compliance decisions. Unlike offline monitors, the proposed online monitor can operate in different scenarios lacking overall information. To adapt to all scenarios and monitor appropriate law articles using little information, this paper proposes the trigger domain-based architecture, where triggers are used to classify environment situations and distinguish the behaviors of a vehicle with the decision information. In addition, the atomic propositions that are necessary for the trigger and judgement MTL expressions are shared in detail for better understanding. The characteristics and challenges in digitizing traffic laws for online monitoring purposes are discussed. The main traffic laws of highways and intersections are taken as digitization examples. A method for calculating fuzzy thresholds in digitized laws based on datasets is introduced, and all the digitized traffic laws are verified on datasets. 

Future work directions could include: (1) a traffic law violation state cancel module behind the judgement part in the trigger domain that considers the around driving conditions to cancel the non-mandatory law violation states for achieving more flexible violation judgement and avoiding subsequent decision-making system having an unsolvable problem; (2) a mapping system based on ontology mapping the atomic propositions to the traffic law nature language, which can ensure that in situations that temporary traffic law changes or driving area changes vehicles can direct access to the violation monitoring MTL expressions without any manual intervention; (3) a traffic law compliance decision-making system that can transfer the MTL expressions into decisions, ensuring full traffic law compliance when safe.

\section*{Acknowledgements}
This work was supported by the National key R\&D Program of China: 2022YFB2503003, and the National Science Foundation of China Project: 52072215 and U1964203.

\bibliographystyle{splncs04}
\bibliography{ref}

\clearpage

\appendix
\setcounter{table}{0}   

\section{Violation Monitor of Article 47}
\label{appendixa}
To facilitate the judgment of the lane change during the overtaking process, the initial lane is considered as a reference lane when generating the overtaking decision, which can be expressed as follows:
\begin{equation}
    \begin{aligned}
    P_{I}(\text{Decision}({Ego}) \neq Overtake) \wedge \text{Decision}({Ego}) = Overtake \Rightarrow \text{InitialLane}(Ego)=\text{Lane}(Ego)\\
    \end{aligned}
\end{equation}
where \textit{I} is the last sampling time of the current time. 

In overtake stage 1, the triggering condition is that the ego vehicle makes the decision to overtake and overlaps with the left lane line, which is given by:
\begin{equation}
    \begin{aligned}
    &T_{overtake1} \Leftrightarrow\\
    &{\text{Decision}({Ego}) = Overtake} \wedge vy({Ego})>0\\ 
    &\wedge \text{overlap}({Area}(Ego),  \text{LaneLine}(\text{InitialLane}(Ego)))\\
    \end{aligned}
\end{equation}

When the triggering condition is satisfied, the stage 1 monitoring module starts monitoring. When the front vehicle is changing lanes to the left, the ego vehicle shall not overtake, which is given by:
\begin{equation}
    \begin{aligned}
    &{FrontnotOvertake} \Leftrightarrow \\
    &\neg \left(\begin{array}{l}
    \text{overlap}({Area}(\emph{Tgt}_{ot}),  \text{LaneLine}(\text{InitialLane}(Ego)))\\ 
    \wedge vy(\emph{Tgt}_{ot})>0 \\
    \end{array}\right)\\
    \end{aligned}
\end{equation}
where $\emph{Tgt}_{ot}$ is the target vehicle to be overtaken. 

It should be noted that $Change\_leftlane\_compliance$ belongs to the changing left lane monitoring module, and it will be called directly when the overtake stage 1 monitoring module starts to monitor.

When the ego vehicle has performed lane-changing to the left, the overtake stage 2 monitoring module starts monitoring, which can be expressed as follows:
\begin{equation}
    \begin{aligned}
    &T_{overtake2}\Leftrightarrow\\
    &{\text{Decision}({Ego}) = Overtake}\wedge \text{Lane}({Ego})=\text{InitialLane}(Ego)-1\\
    &\wedge \neg \text{overlap}({Area}(Ego),  \text{LaneLine}(\text{InitialLane}(Ego)))\\
    \end{aligned}
\end{equation}
In overtake stage 2, according to \textit{Operating specifications for safe and civilised motor vehicle driver--Part2: Requirements for car driving}, the ego vehicle should preferably drive 15 km/h faster than the overtaken vehicle but within the speed limit \cite{GA/T1773}.
In this study, the recommended overtaking speed difference is denoted by $\Delta v_{ot}$. However, when the recommended speed conflicts with the limited speed, the ego vehicle is allowed not to satisfy the recommended speed, which is defined by:
\begin{equation}
    \begin{aligned}
    \emph{RecommendedSpeed} \Leftrightarrow \textit{DiffSpd}(Ego,{\emph{Tgt}_{ot}})>\Delta v_{ot}\\
    \end{aligned}
\end{equation}

When the ego vehicle has passed the target vehicle and begins to overlap with the left lane line of the initial lane, the overtake stage 3 monitoring module starts monitoring, which can be expressed as follows:
\begin{equation}
    \begin{aligned}
    &{overtake3}\Leftrightarrow \\
    &{\text{Decision}({Ego}) = Overtake}\wedge \text{Lane}({Ego})=\text{InitialLane}(Ego)-1\\
    &\wedge \text{overlap}({Area}(Ego),  \text{LaneLine}(\text{InitialLane}(Ego)))\\
    &\wedge \textit{distance}({\emph{Tgt}_{ot}}, Ego)>0 \wedge vy({Ego})<0\\
    \end{aligned}
\end{equation}

The process of returning to the original lane belongs to the behavior of changing lanes to the right, so changing lanes to the right should be monitored, that is, $change\_rightlane\_compliance$. In addition, overtaking from the right is not allowed, so when the ego vehicle makes the decision of overtaking, it is needed to monitor whether the ego vehicle changes lanes to the right, which is given by:
\begin{equation}
    \begin{aligned}
    &T_{overtakeR}\Leftrightarrow\\
    &{\text{Decision}({Ego}) = Overtake}\wedge \text{Lane}({Ego})=\text{InitialLane}(Ego)\\
    \end{aligned}
\end{equation}
\begin{equation}
    \begin{aligned}
    &{OvertakeonRight} \Leftrightarrow \\
    & \text{overlap}({Area}(Ego),  \text{LaneLine}(\text{InitialLane}(Ego)+1)) \wedge vy({Ego})<0\\
    \end{aligned}
\end{equation}

\section{Violation Monitor of Article 38}
\label{appendixb}
The trigger domain and logical judgements of relevant rules considered in this section are presented in Table~\ref{table4}.

\begin{table}[htbp!]
    \footnotesize
	\centering
	\caption{{\normalsize Trigger domain and logic judgements of Article 38 for crossroad}}
	\renewcommand{\arraystretch}{1.0}
	\label{table4}
	\begin{tabular}{p{4cm} p{4cm} p{5cm}}
		\toprule
		Article38 & Trigger domain & Logical requirement \\
		\midrule
		{Traffic light rules} & $\textit{T}_{\textit{TrafficLightRules}}$ & \textit{IllegalPass}=$\bm F$ \\
		{Virtual lane follow rules} & $\textit{T}_{\textit{VirtualLane}}$ & \textit{FollowUsualVirtualLane}=$\bm T$ $\vee \textit{FollowUnusualVirtualLane}$=$\bm T$\\
		\multirow{2}{1.5cm}{Right of way rules} & $\textit{T}_{\textit{RightofWay}}$ & \textit{ViolationRightofWay}=$\bm F$\\
		& $\textit{T}_{\textit{AvoidPedestrian}}$ & \textit{ImpedePedestrian}=$\bm F$\\
		\bottomrule
	\end{tabular}
\end{table}

According to the current traffic light rules, when the red or yellow light turns on, vehicles that have not passed or are not passing the stop line shall stop passing the stop line. That is, if the traffic light is not green both before and after the vehicle crossing the stop line, this vehicle is considered to be violating the traffic light rules.

The trigger domain for the traffic light rule is the time interval between the time the ego vehicle is about to enter the intersection and the time the ego vehicle has entered the intersection, which can be expressed by:

\begin{equation}
    \begin{aligned}
    &{T}_{\emph{TrafficLightRules}}\Leftrightarrow \\
    &(mfrntp(Ego)\in \text{IntersectionArea}) U (mrearp(Ego)\in \text{IntersectionArea})\\
    \end{aligned}
\end{equation}
where \textit{mfrntp}(\textit{obj}) represents the mid-point of the front end; \textit{mrearp}(\textit{obj}) is the mid-point of the back end.

Considering that at the red light, the right-turn vehicles may proceed without interfering with other vehicles and pedestrians, for traffic light rule only, vehicle that decides to turn right will not violate this rule, which is given by:
\begin{equation}
    \begin{aligned}
    &{IllegalPass}\Leftrightarrow\\
    & \neg{\text{Decision}({Ego}) = TurnRight}\\
    &\wedge P_{I}\left(\begin{array}{l}
    Crdn(Ego)\notin \text{IntersectionArea}\\
    \wedge \neg \text{TrafficLight}=\text{G}
    \end{array}\right)\\
    & \wedge (Crdn(Ego)\in \text{IntersectionArea})\\
    & \wedge \neg \text{TrafficLight}=\text{G}\\
    \end{aligned}
\end{equation}
here $I$ is the last sampling time of the current time.

For the virtual lane follow rules, the virtual lane monitoring module is triggered when the ego vehicle is located inside an intersection. In this study, a usual virtual lane is considered the best virtual lane, and an unusual virtual lane is considered a not-recommended virtual lane. The best virtual lane and not-recommended virtual lane are generated based on the initial lane the ego vehicle enters at the intersection and the decision on the moving direction, which is given by:
\begin{equation}
    \begin{aligned}
    &{T}_{VirtualLane}\Leftrightarrow Crdn(Ego)\in \text{IntersectionArea}\\
    \end{aligned}
\end{equation}

The ego vehicle can follow either usual or unusual virtual lane. When the ego vehicle is following the unusual virtual lane, the system will provide a reminder about non-recommendation, which is expressed as follows:
\begin{equation}
    \begin{aligned}
    &\emph{FollowUsualVirtualLane}\Leftrightarrow Crdn(Ego)\in \text{NormVLane}(Ego)\\
    \end{aligned}
\end{equation}
\begin{equation}
    \begin{aligned}
    &\emph{FollowUnusualVirtualLane}\Leftrightarrow \\
    &Crdn(Ego)\in \text{UnuVLane}(Ego)\wedge Crdn(Ego)\notin \text{NormVLane}(Ego)
    \end{aligned}
\end{equation}

For the right-of-way rules, it is necessary to consider avoiding motor vehicles and pedestrians. Pedestrians on sidewalks have the highest right, so when a vehicle passes through the crosswalk, it is necessary to monitor whether there are pedestrians in the crosswalk sub-area or toward the crosswalk sub-area in the adjacent crosswalk sub-area. To avoid motor vehicles with high right-of-way in different moving directions and under different traffic light conditions, different high right-of-way areas have to be considered.

The monitoring module for avoiding motor vehicles with high right-of-way is triggered when the ego vehicle is inside an intersection. When the ego vehicle decides to turn left, vehicles on the opposite go-straight lane have a higher right-of-way. Therefore, if the ego vehicle intends to turn right under the red traffic light, the straight lane on the left side intersects with the ego’s turn-right lane that has a higher right-of-way. The obstructive monitoring area is constructed based on the ego vehicle’s virtual lane and the corresponding lane with a high right-of-way; the existence of the higher right-of-way vehicle is judged based on the direction-judgement line in the opposite lane. To ensure that vehicles with high right-of-way are not affected by the ego vehicle, the ego vehicle should guarantee that there is no high right-of-way vehicle in the obstructive monitoring area when passing the virtual stop line. The corresponding mathematical expressions are as follows:
\begin{equation}
    \begin{aligned}
    {T}_{\emph{RightofWay}}\Leftrightarrow Crdn(Ego)\in \text{IntersectionArea}\\
    \end{aligned}
\end{equation}
\begin{equation}
    \begin{aligned}
    &\emph{ViolationRightofWay}\Leftrightarrow \\
    & \text{overlap}(\text{Line}(mfrntp(Ego), Crdn(Ego)),\text{VStopline}(Ego))\\
    &\wedge \left(\begin{array}{l}
    \exists \emph{Tgt} \in \emph{HighRghtWy}(Ego) \\
    \wedge mfrntp(\emph{Tgt}_{hrw}) \in \text{CheckArea}(Ego)\\
    \end{array}\right)\\
    &\wedge vx(Ego) \neq 0\\
    \end{aligned}
\end{equation}
where $Line(p1, p2)$ indicates the line-connected points $p1$ and $p2$; $\emph{Tgt}_{hrw}$ denotes high right-of-way vehicles under different decisions.
\begin{equation}
    \begin{aligned}
    &{\emph{Tgt} \in \emph{HighRghtWy}(Ego)}\Leftrightarrow \\
    & \left(\begin{array}{l}
    {\text{Decision}({Ego}) = \emph{TurnLeft}}\\
    \wedge \text{overlap}(\text{Area}(\emph{Tgt}), \text{JudgLine}(\emph{Tgt}))\\
    \wedge{Incln}(\emph{Tgt}) \in \text{AngleRange\_GS}\\
    \end{array}\right)
    \vee \left(\begin{array}{l}
    {\text{Decision}({Ego}) = \emph{TurnRight}}\\
    \wedge \neg \text{TrafficLight}=\text{R}\\
    \wedge \text{overlap}(\text{Area}(\emph{Tgt}), \text{JudgLine}(\emph{Tgt}))\\
    \wedge{Incln}(\emph{Tgt}) \in \text {AngleRange\_TL}\\
    \end{array}\right)\\
    &\vee \left(\begin{array}{l}
    {\text{Decision}({Ego}) = \emph{TurnRight}}\\
    \wedge \text{TrafficLight}=\text{R}\\
    \wedge \text{overlap}(\text{Area}(\emph{Tgt}), \text{JudgLine}(\emph{Tgt}))\\
    \wedge{ Incln}(\emph{Tgt}) \in \text {AngleRange\_GS}\\
    \end{array}\right)\\
    \end{aligned}
\end{equation}
where $\text{Check\_area}$ is the obstructive monitoring area concerning $\emph{Tgt}_{hrw}$; ${Incln}(obj)$ is the deviation angle between the \textit{obj} and the centre line of the oncoming lane when the vehicle crosses the direction-judgement line, with a positive angle for the counterclockwise direction; $\text{AngleRange\_GS}$ and $\text{AngleRange\_TL}$ denote the heading angle ranges when the vehicle goes straight or turns left, respectively.

As mentioned above, when the ego vehicle enters the pedestrian crossing, to avoid pedestrians, there should be no pedestrian in the corresponding crosswalk sub-area. Meanwhile, there should be no pedestrians heading toward this area in the adjacent crosswalk sub-area, which can be expressed as follows:
\begin{equation}
    \begin{aligned}
    &{T}_{AvoidPedestrian}\Leftrightarrow\\
    & mfrntp(Ego) \in \text{CWSubA}(Ego)\wedge Crdn(Ego) \notin \text{CWSubA}(Ego)\\
    \end{aligned}
\end{equation}
\begin{equation}
    \begin{aligned}
    &{ImpedePedestrian}\Leftrightarrow\\
    &\exists Crdn(ped) \in \text{CWSubA}(Ego)
    \vee \left(\begin{array}{l}
    Crdn(ped) \in \text{CWSubAdj}(Ego)\\
    \wedge \text{direction}(ped, \text{CWSubA}(Ego)) \\
    \end{array}\right)\\
    \end{aligned}
\end{equation}

In the above expressions, pedestrians are regarded as points with a heading direction, and $ped$ denotes the coordinates of the pedestrians; $\text {CWSubAdj}$ indicates the adjacent sub-area of the crosswalk sub-area where the ego vehicle is located in; $\text{direction}(ped, \text{CWSubA}(Ego))$ represents the direction heading towards the subarea where the ego vehicle is located in.

\section{Atomic Proposition}
\label{appendixc}
\begin{table}\centering
\footnotesize
\caption{Concepts involved in atomic propositions}
\label{AP}
\begin{tabular}{p{3cm} p{13cm}}
    \toprule
    \multicolumn{1}{c}{\textbf{Label}}&\multicolumn{1}{c}{\textbf{Meaning}} \\ \midrule
    \multicolumn{1}{c}{\textbf{Object}} &\\
    \textit{Ego} & The structure of ego vehicle \\
    \textit{Tgt} & The structure of other traffic participant\\
    \textit{Ped} & The structure of pedestrian\\
    $\emph{Tgt}_{f}$ & The \textit{Tgt} that located in the front area of \textit{Ego}, that is, Area$( \emph{Tgt}_{f}) \wedge \text{Front}({Ego}) \neq \varnothing$ \\
    $\emph{Tgt}_{fl}$,$\emph{Tgt}_{fr}$,$\emph{Tgt}_{r}$, $\emph{Tgt}_{rl}$,$\emph{Tgt}_{rr}$ & Similar to $\emph{Tgt}_{f}$, represent the \textit{Tgt} located in the rest five regions of \textit{Ego}\\
    Line(\textit{points}) & The line segment composed of two points\\
    LaneLine(\textit{i}) & Cubic fitting curve of \textit{i}th lane line \\
    IntersectionArea & The structure of the area that enclosed by stop line extensions\\
    TrafficLight & The traffic light status. R,G,Y represents the red, green and yellow light respectively\\
    \multicolumn{1}{c}{\textbf{Property}} &\\
    \textit{vx}(\textit{obj}) &  Obtain longitudinal speed of \textit{obj}\\
    \textit{vy}(\textit{obj}) &  Obtain lateral speed of \textit{obj}\\
    $\theta$(\textit{obj}) & Obtain heading angle of \textit{obj}\\
    \textit{w}(\textit{obj}) & Obtain width of \textit{obj}\\
    \textit{l}(\textit{obj}) & Obtain length of \textit{obj}\\
   
    \textit{mfrntp}(\textit{obj}) & Obtain mid-point coordinate of the front end of \textit{obj}\\
    \textit{mrearp}(\textit{obj}) & Obtain mid-point coordinate of the rear end of \textit{obj}\\
    \textit{Area}(\textit{obj}) & Obtain the area structure that include the vertex coordinate structure of \textit{obj}\\
    \textit{Crdn}(\textit{obj}) & Obtain geometric center coordinate of \textit{obj}\\
    Decision(\textit{obj}) & Obtain vehicle decision information structure\\
    Lane(\textit{obj}) & Obtain the lane order number which \textit{obj} belongs to\\
    $\text{RoadID}_{in}$(\textit{obj}) & Obtain the road ID when \textit{obj} enters the intersection\\
    $\text{RoadID}_{out}$(\textit{obj}) & Obtain the road ID when \textit{obj} leaves the intersection\\
    RoadType(\textit{obj}) & Obtain road type code of \textit{obj}, M, R, A and D represents the mainway, ramp, acceleration lane and deceleration lane respectively\\
    InitialLane(\textit{obj}) & Obtain the lane of \textit{obj} when \textit{obj} decides to overtake\\
    NormVLane(\textit{obj}) & Obtain \textit{obj}’s best driving area to pass through the intersection\\
    UnuVLane(\textit{obj}) & Obtain \textit{obj}’s compliance but not recommended area to  pass through the intersection\\
    VStopline(\textit{obj}) & Obtain the stop line segment that \textit{obj} should stop before it to prevent interfering \textit{Tgt} with high right-of-way\\
    CheckArea(\textit{obj}) & Obtain the area used to judge whether there is \textit{Tgt} with higher right-of-way than \textit{obj}\\
    JudgLine(\textit{obj}) & Obtain the line segment used to judge the driving direction of \textit{obj}\\
    CWSubA(\textit{obj}) & Obtain the square area with side length equals to lane width in the crosswalk where \textit{obj} is located in \\
    CWSubAdj(\textit{obj}) & Obtain the adjacent square area of  CWSubA(\textit{obj})\\
     Front(\textit{obj}) & Obtain the ahead front area structure of \textit{obj}\\
    FrontLeft(\textit{obj}) & Obtain the front left area structure of \textit{obj}\\
    FrontRight(\textit{obj}) & Obtain the front right area structure of \textit{obj}\\
    Rear(\textit{obj}) & Obtain the rear area structure of \textit{obj}\\
    RearLeft(\textit{obj}) & Obtain the rear left area structure of \textit{obj}\\
    RearRight(\textit{obj}) & Obtain the rear right area structure of \textit{obj}\\
    \multicolumn{1}{c}{\textbf{Relationship}} &\\
    SpdSignArea(\textit{obj}) & Judge whether \textit{obj} is located in the speed limit sign management area,\\
    & $s(speedsign)<s(obj)<s(lift\_speedsign)$\\
    Overlap(\textit{obj}1,\textit{obj}2) & Judge the overlap status of \textit{obj}1 and \textit{obj}2, \textit{obj}1 $\wedge$ \textit{obj}2 $\neq \varnothing$\\
    LngTmOnLine(\textit{obj}) & Judge whether \textit{obj} overlap with lane line longer than the maximum allowable time,\\ 
    & \textit{t}$($Overlap$($\textit{obj}$,$LaneLine$($Lane$($\textit{obj}$))))$ $\neq \varnothing$\\
    HighRghtWy(\textit{obj}) & Judge whether exist vehicle that has higher right-of-way than \textit{obj} in the current scene\\
    Direction(\textit{obj},$Area$) & Judge whether the direction of \textit{obj} points to the area.\\
    \multicolumn{1}{c}{\textbf{Calculate}} &\\
     \textit{s}(\textit{obj}1, \textit{obj}2) &  Calculate the longitudinal distance between  \textit{obj}1 and \textit{obj}2\\
     \textit{distance}(\textit{obj}1, \textit{obj}2) & Calculate the longitudinal distance of \textit{obj}1 and \textit{obj}2, \textit{obj}2 is in front of \textit{obj}1 \\
    \textit{TTCX}(\textit{obj}1,\textit{obj}2) & Calculate the time to longitudinal collision of \textit{obj}1 and \textit{obj}2, \textit{obj}2 is in front of \textit{obj}1\\
    \textit{DiffSpd}(\textit{obj}1,\textit{obj}2) & Calculate the speed difference between \textit{obj}1 and \textit{obj}2\\
    \textit{Incln}(\textit{obj}) & Calculate the deviation angle between \textit{obj} and the road centre line when \textit{obj} entering the intersection.\\
    \bottomrule
\end{tabular}
\end{table}
%




\end{document}